\documentclass{article}
\usepackage[utf8]{inputenc}
\usepackage[margin=1in]{geometry}
\usepackage{graphicx}
\usepackage{subcaption}
\usepackage{amsmath}
\numberwithin{equation}{section}
\usepackage{booktabs}
\usepackage{amssymb}
\usepackage[dvipsnames]{xcolor}
\usepackage{todonotes}
\usepackage{tcolorbox}
\usepackage{tikz}
\usetikzlibrary{matrix,chains,decorations.pathreplacing,arrows,fit,positioning,calc}
\usepackage{lscape}
\usepackage{tabularx,tabulary,array}
\newcolumntype{M}[1]{>{\centering\arraybackslash}m{#1}}
\usepackage{authblk}

\definecolor{HCrimson}{RGB}{165,28,48}
\definecolor{HInk}{RGB}{30,30,30}
\definecolor{HMortar}{RGB}{140,129,121}
\definecolor{HParchment}{RGB}{243,243,241}
\definecolor{HSlate}{RGB}{137,150,160}
\definecolor{HShade}{RGB}{186,197,198}
\definecolor{HIvy}{RGB}{82,133,76}
\definecolor{HSafron}{RGB}{232,125,30}

\newcommand{\lr}[1]{\left(#1\right)}
\newcommand{\vel}{\mathbf{u}} 
\newcommand{\press}{p} 
\newcommand{\lapl}{\nabla^{2}} 
\renewcommand{\time}{\mathrm{t}}
\newcommand{\vave}{\overline{\vel}} 
\newcommand{\uave}{\overline{u}} 
\newcommand{\pave}{\overline{\press}} 
\newcommand{\vfluct}{\vel^{\prime}} 
\newcommand{\pfluct}{\press^{\prime}} 
\newcommand{\Rey}{\mathrm{Re}} 
\newcommand{\rs}{\boldsymbol{\mathcal{R}}} 
\newcommand{\anirs}{\mathbf{a}} 
\newcommand{\nanirs}{\mathbf{b}} 

\newcommand{\Shat}{\widehat{\mathbf{S}}}
\newcommand{\Rhat}{\widehat{\mathbf{R}}}

\usepackage{hyperref}
\hypersetup{
    colorlinks=true,
    linkcolor=blue,
    filecolor=magenta,      
    urlcolor=cyan,
}

\usepackage[
backend=biber,
style=ieee,
sorting=nyt
]{biblatex}
\title{Deep learning for turbulent channel flow} 
\author[1]{Rui Fang}
\author[1]{David Sondak}
\author[1]{Pavlos Protopapas}
\author[1,2]{Sauro Succi}
\affil[1]{Institute for Applied Computational Science, Harvard University, 33 Oxford Street, Cambridge, MA 02138, USA}
\affil[2]{Center for Life Nanoscience, Italian Institute of Technology, Viale Regina Margherita 295, 00161, Roma, Italy}

\date{}

\begin{document}

 \maketitle

\begin{abstract}
Turbulence modeling is a classical approach to address the multiscale nature of fluid turbulence.  Instead of resolving all scales of motion, which is currently mathematically and numerically intractable, reduced models that capture the large-scale behavior are derived.  One of the most popular reduced models is the Reynolds averaged Navier-Stokes (RANS) equations.  The goal is to solve the RANS equations for the mean velocity and pressure field.  However, the RANS equations contain a term called the Reynolds stress tensor, which is not known in terms of the mean velocity field.  Many RANS turbulence models have been proposed to model the Reynolds stress tensor in terms of the mean velocity field, but are usually not suitably general for all flow fields of interest.  Data-driven turbulence models have recently garnered considerable attention and have been rapidly developed.  In a seminal work, Ling et al (2016) developed the tensor basis neural network (TBNN), which was used to learn a general Galilean invariant model for the Reynolds stress tensor.  The TBNN was applied to a variety of flow fields with encouraging results.  In the present study, the TBNN is applied to the turbulent channel flow.  Its performance is compared with classical turbulence models as well as a neural network model that does not preserve Galilean invariance.  A sensitivity study on the TBNN reveals that the network attempts to adjust to the dataset, but is limited by the mathematical form that guarantees Galilean invariance.

\end{abstract}

\section{Introduction} \label{sec:intro}
Fluids touch every aspect of life on Earth and exhibit a wide range of complex and wonderful behavior.  A prime objective of
science and engineering is to understand the forces a fluid exhibits on its surroundings.  Understanding such forces is a
necessity for predictive science and engineering.  For example, an accurate description of the forces on an airplane wing can
help engineers design more efficient and cost-effective airplanes.  Predicting the pressure that blood exerts on the walls of
arteries has important implications for medical science and health.  Outside of Earth, man-made satellites are prone to the
solar wind; accurate predictions of space weather will have a significant impact on satellite systems.  The fluid systems
just described cover an exceptionally broad range of fluid mechanics from aerodynamics through biological fluid flows to
space plasmas.  All of these systems require models for fluid behavior usually coupled with constitutive models to describe
physical processes at smaller scales not included in the primary fluid models.  The main continuum fluid model is given by
the Navier-Stokes equations, which expresses conservation of mass and momentum.  The goal is to solve the Navier-Stokes
equations for the velocity field from which fluid forces can be derived.  Additional physical processes can be included by
modifying the stress tensor as well as including body forces in the momentum equation.  However, even in their simplest form,
the Navier-Stokes equations are formidable and pose a significant challenge to scientists, engineers, and mathematicians.

Although the Navier-Stokes equations have been known for more than 150 years, exact mathematical solutions are extremely
rare.  This is a problem because the quantities of interest, such as the drag force, are determined from the velocity field
obtained by solving the Navier-Stokes equations.  In the absence of mathematical techniques, one could employ computational
approaches to numerically solve the Navier-Stokes equations and determine the velocity field.  This is precisely the approach taken in 
direct numerical simulations (DNS) in which the governing equations are discretized and solved directly on a computer.  In order to 
achieve accurate solutions, all scales of motion must be resolved.  This is an enormous challenge and generally makes DNS intractable 
for practical problems~\cite{pope2001turbulent}.  The reason for these challenges is that the Navier-Stokes equations give rise to the 
physical phenomenon of turbulence, which is the result of the underlying nonlinearity in the governing equations.  This nonlinearity is 
responsible for multiple interacting spatial and temporal scales.  Turbulent flow appears to be the rule rather than the exception; most 
fluid flows found in nature are in a state of turbulence.  In order to design better systems, one must be able to
predict the overall behavior of a turbulent flow field.  In fact, scientists and engineers are primarily interested in the
large scale behavior of a turbulent flow field rather than the exact detailed dynamics.  Accurate prediction of the largest scales
often leads to acceptable predictions of quantities of interest.  The field of turbulence modeling is concerned with the
development of models that account for the effects of the smaller scales of motion on the large scales of 
motion~\cite{wilcox1998turbulence}.  Through this route, practitioners seek to develop reduced models that are more mathematically and 
computationally accessible.

Although no rigorous definition of turbulence exists, the phenomenon does exhibit some well-accepted 
features~\cite{batchelor1953theory, tennekes1972first, pope2001turbulent}.  Visually, a turbulent flow field appears as almost complete 
disorder.  What order there is, is manifested as intermittent coherent structures at different spatial and temporal scales.  For example, 
large and small whirls of fluid interact with each other, sometimes combining with each other and other times destroying each other.  At 
its heart, turbulence is a multiscale phenomenon without any scale separation.  That is, there is a continuum of scales in a turbulent flow 
field all interacting with each other none of which can be out-right neglected.  This fact has made turbulence an exceptionally difficult 
problem to model; there is no obvious scale-separation at which one could introduce a model for turbulence.  

One of the earliest attempts at modeling turbulence, due to Osborn Reynolds, is to decompose the velocity field into an average and 
fluctuations about the average~\cite{reynolds1895dynamical}. This split is known as the Reynolds-averaged decomposition, the idea being 
that the average behavior of the flow field is sufficient to determine quantities of interest.  The Reynolds decomposition is an example of 
a coarse-graining operation, in which the governing equations, containing all possible dynamics, are reduced to a set of equations for only 
the scales of motion that are of interest.  Applying the Reynolds decomposition to the Navier-Stokes equations results in the 
Reynolds-Averaged Navier Stokes (RANS) equations, which are to be solved for the average velocity field.  However, the RANS equations 
contain a new term, called the Reynolds stress tensor, whose mathematical form does not include an explicit dependence on the average 
velocity field.  The effect of the Reynolds stress tensor on the average velocity field is of paramount importance and has been the focus 
of turbulence modeling for many decades.  A variety of models have been introduced with varying degrees of sophistication and 
success~\cite{wilcox1998turbulence}.  Popular models include the famous eddy-viscosity models in which the Reynolds stress tensor is 
proportional the gradients of the average velocity field~\cite{boussinesq1877theorie, schmitt2007boussinesq}.  This proportionality is 
expressed through an eddy viscosity, which accounts for momentum transport by the turbulent eddies.  Generalizations 
to the standard eddy viscosity approach have included dependence on powers of the velocity gradients to account for more physical effects 
such as recirculation regions~\cite{craft1996development, pope1975more}.  Researchers have also attempted to include non-local effects into 
eddy-viscosity models for the Reynolds stress tensor by relating it to the time-history of the velocity 
gradients~\cite{hamba2005nonlocal, hamlington2008reynolds}.  Nonlocal effects can also be included by deriving additional 
transport equations for the Reynolds stress tensor~\cite{pope2001turbulent}.  These new equations include other terms that must be 
modeled.  Such sophisticated models tend to shed one of the primary advantages of eddy-viscosity models: their ease of implementation.  
Over the years, standard models have been implemented in most engineering fluid dynamics software packages.  Unfortunately, different 
models apply to different flow fields, which is certainly not ideal.  Moreover, even within a single model, there can be multiple tunable 
parameters to tweak in order to get good agreement with different flow fields.  For example, wall-bounded flows (such as a duct-flow) may 
use a different model than free-surface flows (such as flow around an airfoil).  Given these challenges, researchers and modelers have 
recognized the potential offered by including data from experiments or numerical simulations into turbulence models.

In recent years, data science has been leveraged in a number of fields to tackle challenging problems~\cite{lecun2015deep} including 
natural language processing~\cite{collobert2008unified} and speech and image recognition~\cite{nasrabadi2007pattern}.  Data science has 
already had a transformative impact in the business world~\cite{waller2013data} and has been impactful in the tech and finance industries.  
Very recently, scientists and engineers have started to explore and adapt techniques and algorithms from the traditional data science 
community to scientific problems\cite{carrasquilla2017machine, baldi2014searching}. Machine learning and other techniques from data science 
have started to percolate through scientific fields as diverse as DNA sequencing~\cite{libbrecht2015machine} and 
the discovery of new materials~\cite{pilania2013accelerating}.  Fluid mechanics researchers have now started to actively develop data 
science techniques for fluid mechanics systems~\cite{brunton2016discovering, jimenez2018machine}.

One of the first problems to be tackled by machine learning algorithms in fluid mechanics was how to learn a good model for the Reynolds 
stress tensor and considerable effort has been devoted to this task~\cite{tracey2015machine, zhang2015machine, wang2017physics}.  Recently, 
researchers have used random forests and neural networks to learn adequate models for the Reynolds stress 
tensor~\cite{ling2016machine, ling2016reynolds, wu2018physics}.  Most machine learning algorithms for learning Reynolds stress models have 
been supervised.  This means that they learn the model with data from direct numerical simulations (DNS).  The DNS data is generated from 
high-fidelity physical models that express non-negotiable conservation laws such as conservation of mass and momentum.  However, without 
regulation a machine learning algorithm can fit a model to the DNS data in any way that it deems appropriate, even if that means a 
violation of the governing physical laws.  Therefore, there is considerable interest in developing physics-aware machine learning 
algorithms.  In a seminal work, Ling et al.~\cite{ling2016reynolds} proposed a neural network architecture, called the tensor basis neural 
network (TBNN), that learns a model for the Reynolds stress tensor that is exactly Galilean-invariant.  The TBNN model was trained on a 
variety of flow fields including a turbulent channel flow, flow over a backward facing step, and flow around a square cylinder, among 
others.  When tested on flow over a wavy wall, the TBNN results showed considerable improvement over a standard neural network architecture.

In spite of their apparent success, there are relatively few studies that elucidate the actual learning process inside a neural network.  
The present work focuses on the TBNN architecture and analyzes its performance on the turbulent channel flow.  In particular, the specific 
predictions from the TBNN of relevant components of the Reynolds stress are assessed.  These predictions are compared to the classical 
linear and quadratic eddy viscosity models as well as a fully-connected neural network that is not aware of Galilean invariance.  The  
remainder of the paper is structured as follows.  Section~\ref{sec:bg} provides background on the equations of fluid mechanics, 
turbulence modeling, channel flow, and neural networks.  Following this, the methodology used to train the network is discussed in 
Section~\ref{sec:methods} including specific simulation parameters.  Results are presented and discussed in Section~\ref{sec:results}.  
Finally, conclusions, limitations of the present study, and future work are described in Section~\ref{sec:conclusions}.

\section{Background} \label{sec:bg}
Scientists and engineers seek to predict and understand the velocity $\vel$ and pressure $\press$ fields of a variety of fluid flows.  The 
challenge facing scientists and engineers is to solve the incompressible Navier-Stokes equations,
\begin{align}
  \frac{\partial \vel}{\partial \time} + \nabla\cdot\left(\vel\otimes\vel\right) &= -\frac{1}{\rho}\nabla \press + \nu\lapl\vel \label{eq:mom} \\
  \nabla\cdot\vel &= 0 \label{eq:cont}
\end{align}
for the three-dimensional velocity field $\vel = \lr{u, v, w}$ and the pressure field $\press$, which are functions of space and time.  The parameters $\nu$ and $\rho$ are the kinematic viscosity and density of the fluid, respectively.  The key dimensionless parameter in incompressible fluid mechanics, the Reynolds number $\Rey$, is formed by a velocity scale $U$ and a length scale $L$ and is given by $\Rey = UL/\nu$.  A large $\Rey$ indicates that the fluid flow is turbulent whereas a small $\Rey$ suggests a laminar flow field.  Many flows of scientific and engineering interest are in a turbulent regime, which is characterized by many simultaneously active temporal and spatial scales.  Analytical approaches to solving the Navier-Stokes equations have succeeded for only the simplest flow fields, usually in idealised geometries.  Numerical approaches for resolving turbulent flow fields are strained by the multiscale nature of turbulence and are ultimately restricted to relatively low $\Rey$ flows, currently around $10^4 - 10^5$. This is in contrast to a standard automobile, which features $\Rey \sim 10^7$.

\subsection{Turbulence modelling}\label{sec:TM}
The earliest rigorous mathematical attempt at resolving the turbulence problem was due to Osborn Reynolds~\cite{reynolds1895dynamical}.  The 
idea is to decompose the fields into average and fluctuating components,  
\begin{align}
  \vel &= \vave + \vfluct \label{eq:vdecomp} \\
  \press &= \pave + \pfluct \label{eq:pdecomp}
\end{align}
where $\overline{\lr{\cdot}}$ denotes and averaged quantity and $\lr{\cdot}^{\prime}$ denotes a fluctuating component.  
Introducing~\eqref{eq:vdecomp} and~\eqref{eq:pdecomp} into the Navier-Stokes equations and averaging results in the Reynolds averaged 
Navier-Stokes (RANS) equations, 
\begin{align}
  \frac{\partial \vave}{\partial \time} + \nabla\cdot\left(\vave\otimes\vave\right) &= -\frac{1}{\rho}\nabla\pave + \nu\lapl\vave -
\nabla\cdot\rs \label{eq:momave} \\
  \nabla\cdot\vave &= 0. \label{eq:contave}
\end{align}
where 
\begin{align}
  \rs = \overline{\vfluct\otimes\vfluct} \label{eq:rstress}
\end{align}
is called the Reynolds stress tensor.  With the notable exception of the Reynolds stress tensor, the RANS equations 
are identical to the Navier-Stokes equations.  The core challenge of this approach is that the Reynolds stress tensor 
is unknown.  That is, the evolution equation for the first moment of the velocity field (the average, $\vave$) depends 
upon the second moment of the velocity field (the covariance).  Additional relations must be specified to determine 
$\rs$ and close the system of equations.  Although transport equations can be derived for the Reynolds stresses, these involve third-order 
moments of the velocity field.  Indeed, attempting to close the RANS equations results in an infinite cascade of unclosed terms.  Efforts 
have therefore focused primarily on modeling the effects of the Reynolds stress tensor on the average fields.  The goal of turbulence modeling is to propose useful and tractable models for $\rs$.

Note that the Navier-Stokes equations have a variety of transformation properties.  Of particular consequence in the 
present work is Galilean invariance.  That is, the Navier-Stokes equations are the same in an inertial reference frame 
that is translating with a constant velocity $\mathbf{V}$.  Hence, replacing the spatial coordinate with 
$\mathbf{x} - \mathbf{V}t$ and the velocity with $\mathbf{U} - \mathbf{V}$ does not change the form of the 
Navier-Stokes equations.  This fact remains true even for the RANS equations.  Therefore, any turbulence model for 
the Reynolds stress tensor must also be Galilean invariant.

\subsubsection{Reynolds stress tensor}\label{sec:reynolds_stress_tensor}
The Reynolds stress tensor has been studied extensively and many properties are known regarding its structure.  It is a symmetric, second order tensor with known invariants~\cite{pope2001turbulent}.  The anisotropic component of $\rs$ is responsible for turbulent transport and therefore modelling efforts have focused on the anisotropic Reynolds stress tensor,  
\begin{align}
  \anirs = \overline{\vfluct\otimes\vfluct} -\frac{2}{3}k\mathbf{I},
  \label{eq:anisotropy}
\end{align}
where the turbulent kinetic energy $k\left(\mathbf{x},t\right)$ is  given by   
\begin{align*}
  k = \frac{1}{2} \overline{\vfluct \cdot \vfluct} = 
      \frac{1}{2} \mathrm{trace}\lr{\overline{\vfluct \otimes \vfluct }}
\end{align*}
and $\mathbf{I}$ is the identity matrix.  In this work, we will be concerned with the normalized anisotropy tensor, 
\begin{align}
  \nanirs = \frac{\anirs}{2k}.
  \label{eq:normalized_anisotropy}
\end{align}
We indicate individual components of the tensor with subscripts, corresponding to which velocity correlations are involved. For example, $b_{uv} = \overline{uv} / 2k$.

\subsubsection{Linear eddy-viscosity model}\label{sec:LEVM}
Significant modelling efforts have been devoted to finding closures for the Reynolds stresses~\cite{jones1972prediction, pope1975more, craft1996development}.  
A very popular approach is to represent the Reynolds stresses using a so-called eddy viscosity,
\begin{align}
  \anirs = -2\nu_{T}\overline{\mathbf{S}} \label{eq:levm}
\end{align}
where $\overline{\mathbf{S}} = \frac{1}{2}\left(\nabla\vave + \left(\nabla\vave\right)^{\mathsf{T}}\right)$ is the mean rate of 
strain tensor and $\nu_{T}$ is called the eddy viscosity.  The model given by~\eqref{eq:levm} is called the linear eddy viscosity model 
(LEVM) because the Reynolds stresses are a linear function of the mean velocity gradients.  The eddy viscosity model is motivated via 
analogy with the molecular theory of gases.  The turbulent flow is thought of as consisting of multiple interacting eddies.  The eddies 
exchange momentum giving rise to an eddy viscosity.  Although convenient, the eddy viscosity hypothesis is known to be incorrect for many 
flow fields.  The intrinsic assumption that the Reynolds stresses only depend on local mean velocity gradients is incorrect; turbulence is 
a temporally and spatially nonlocal phenomenon.
Moreover, the specific form proposed in analogy with the molecular theory of gases~\eqref{eq:levm} is also flawed because the turbulence timescales are at odds with the timescales in the molecular theory of gases.  Nevertheless, the eddy viscosity model is appealing due to its simplicity and ease of numerical implementation.

A form for the eddy viscosity $\nu_{T}$ must be specified to complete the LEVM given by~\eqref{eq:levm}.  One of the most commonly 
used forms for the eddy viscosity is the $k-\epsilon$ model~\cite{jones1972prediction},
\begin{align}
  \nu_{T} = C_{\mu}\frac{k^{2}}{\epsilon} 
  \label{eq:keps}
\end{align}
where $\epsilon$ is the turbulent dissipation.  In general, the model constant $C_{\mu}$ must be calibrated for different flows.  A common choice is $C_{\mu} = 0.09$, which has been observed in channel flow~\cite{kim1987turbulence, pope2001turbulent} and the temporal mixing layer~\cite{rogers1994direct, pope2001turbulent}. In fact, for simple shear flows, empirical evidence suggests that this is the correct value for $C_{\mu}$~\cite{pope2001turbulent}. Finally, transport equations for $k$ and $\epsilon$ are solved along with the RANS equations.
In terms of the linear eddy viscosity model, the RANS equations are,
\begin{align}
  \frac{\partial \vave}{\partial \time} + \nabla\cdot\lr{\vave\otimes\vave} &= -\frac{1}{\rho}\nabla\pave + \nabla\cdot\lr{\lr{\nu + \nu_{T}}\lr{\nabla\vave + \nabla\vave^{\mathsf{T}}}}
\end{align}
where $\nu_{T}$ is determined from~\eqref{eq:keps}.  The turbulent kinetic energy $k$ and dissipation $\epsilon$ are solved from their respective transport equations, which we omit here for brevity.
Expressed in terms of the normalized anisotropy tensor~\eqref{eq:normalized_anisotropy}, the $k-\epsilon$ model becomes,
\begin{align}
    \nanirs = -C_{\mu} \Shat
    \label{eq:levm_normalized}
\end{align} 
where $\Shat=\frac{k}{\epsilon} \overline{\mathbf{S}}$ is the normalized mean rate of strain tensor. 

The deficiencies of the $k-\epsilon$ model and the isotropic eddy viscosity assumption have been well-documented: namely, the inability to 
account for streamline curvature and history effects \cite{speziale1991analytical, gatski2004constitutive, johansson2002engineering}.   
Nonlinear eddy viscosity models, although more computationally expensive, have the potential to represent additional flow physics, such as 
secondary flows and flows with mean streamline curvature.  Nonlinear models have been developed including quadratic eddy viscosity models.  In the present work, we compare results with a specific quadratic eddy viscosity model~\cite{craft1996development}.  In the next section, we describe a general nonlinear eddy viscosity model.

\subsubsection{General eddy viscosity model}\label{sec:GEVM}
The most general representation of the anisotropic Reynolds stresses in terms of the mean rate of strain and rotation is~\cite{pope1975more},
\begin{align}
  \nanirs = \sum_{n=1}^{10}g^{(n)}\left(\lambda_1, ..., \lambda_5\right) \mathbf{T}^{(n)}
  \label{eq:nlev}
\end{align}
where $\mathbf{T}^{(n)}$ are tensors depending on the normalized rate of strain and rotation.  The form~\eqref{eq:nlev} guarantees Galilean 
invariance and ensures that predictions made with this model are not dependent on the orientation of the coordinate axes.  If this were not 
satisfied, then fluid behavior would be different for observers in different frames of reference.  In order to achieve the desired 
invariance, the coefficients of the tensor basis must depend on the five scalar tensor invariants $\lambda_m, \quad m = 1,\ldots,5$. The 
basis tensors and the invariants are known functions of the normalized mean rate of strain and rotation, $\Shat$ and $\Rhat$, respectively, and are given by,
     \begin{equation}
         \begin{split}
             \mathbf{T}^{(1)} &= \Shat \\
             \mathbf{T}^{(2)} &= \Shat\Rhat - \Rhat\Shat \\
             \mathbf{T}^{(3)} &= \Shat^{2} - \frac{1}{3}\mathrm{Tr}\left(\Shat^{2}\right)\mathbf{I} \\
             \mathbf{T}^{(4)} &= \Rhat^{2} - \frac{1}{3}\mathrm{Tr}\left(\Rhat^{2}\right)\mathbf{I} \\
             \mathbf{T}^{(5)} &= \Rhat\Shat^{2} - \Shat^{2}\Rhat  \\
         \end{split}
         \quad
         \begin{split}
             &\mathbf{T}^{(6)}  = \Rhat^{2}\Shat + \Shat\Rhat^{2} - \frac{2}{3}\mathrm{Tr}\left(\Shat\Rhat^{2}\right)\mathbf{I} \\
             &\mathbf{T}^{(7)}  = \Rhat\Shat\Rhat^{2} - \Rhat^{2}\Shat\Rhat \\
             &\mathbf{T}^{(8)}  = \Shat\Rhat\Shat^{2} - \Shat^{2}\Rhat\Shat \\
             &\mathbf{T}^{(9)}  = \Rhat^{2}\Shat^{2} + \Shat^{2}\Rhat^{2} - \frac{2}{3}\mathrm{Tr}\left(\Shat^{2}\Rhat^{2}\right)\mathbf{I} \\
             &\mathbf{T}^{(10)} = \Rhat\Shat^{2}\Rhat^{2} - \Rhat^{2}\Shat^{2}\Rhat
         \end{split}
         \label{eq:basistensors}
     \end{equation}
     where
     \begin{equation}
        \begin{split}
            \Shat &= \frac{k}{2\epsilon} \left(\nabla \vave + \left(\nabla \vave\right)^{\mathsf{T}}\right) \\
            \Rhat &= \frac{k}{2\epsilon} \left(\nabla \vave - \left(\nabla \vave\right)^{\mathsf{T}}\right)
        \end{split}
        \label{eq:nstrainrotation}
     \end{equation}
The invariants are, 
     \begin{equation}
         \lambda_1 = \mathrm{Tr}\left(\Shat^2\right), \quad
         \lambda_2 = \mathrm{Tr}\left(\Rhat^2\right), \quad
         \lambda_3 = \mathrm{Tr}\left(\Shat^3\right), \quad
         \lambda_4 = \mathrm{Tr}\left(\Rhat^2\Shat\right), \quad
         \lambda_5 = \mathrm{Tr}\left(\Rhat^2\Shat^2\right).
         \label{eq:invariants}
     \end{equation}
Note that the linear eddy viscosity model is recovered when $g^{\left(1\right)} = -C_{\mu}$ and $g^{\left(n\right)} = 0$ for $n>1$.

Finding the coefficients of~\eqref{eq:nlev} is extremely difficult for general three-dimensional turbulent flows, with the aggravation that 
there is no obvious hierarchy of the basis components.  There are additional shortcomings of the representation of $\mathbf{b}$ 
via~\eqref{eq:nlev} beyond its obvious complexity.  For example, the Reynolds stresses are not necessarily functions solely of the mean 
rate of strain and rotation.  Building on this point, the Reynolds stresses are nonlocal objects and representing them as functions of 
local quantities is insufficient.  Nevertheless, the representation~\eqref{eq:nlev} for the eddy viscosity is appealing because the tensor 
basis is an integrity bases which guarantees that $\mathbf{b}$ will satisfy Galilean invariance and remain a symmetric, anisotropic tensor \cite{pope1975more}.

Although~\eqref{eq:nlev} is very general, it is also very complicated.  The approach taken in~\cite{ling2016reynolds} was to train a deep neural network architecture to learn the tensor basis coefficients and subsequently the Reynolds stress tensor across a variety of flow fields.  In the next section, we briefly review the canonical flow field that is the subject of this paper.

\subsection{Physics of turbulent channel flow} \label{sec:channelflow}
\begin{figure}[h!]
    \centering
    \includegraphics[width=0.6\textwidth]{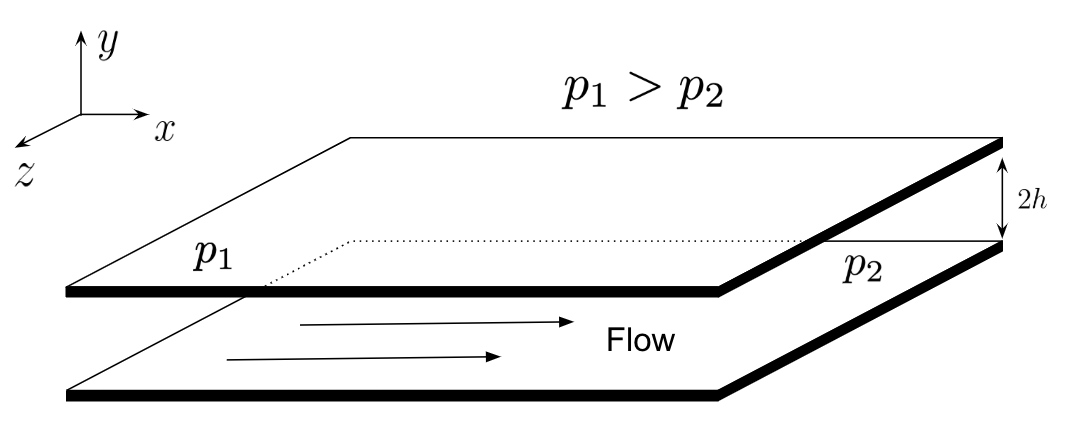}
    \caption{Channel flow geometry. The pressure at the channel inlet $p_1$ is higher than the pressure at the channel exit $p_2$. }
    \label{fig:channelflow}
\end{figure}

We briefly review a few key concepts of the physics of turbulent channel flow.  Additional details can be found  in~\cite{pope2001turbulent}.  Turbulent channel flow is a pressure-driven flow between two parallel planes (see figure~\ref{fig:channelflow}).  The planes are located at $y=-h$ and $y=h$ and the flow proceeds primarily along the $x-$direction.  The direction normal to the wall is the $y-$direction.  Fully-developed, turbulent channel flow shows a one-dimensional structure along the $y$ direction.  That is, after performing the averaging procedure, the flow quantities (such as average velocity) are only functions of the distance across the channel, $y$.  

Using the fact that the turbulent channel flow is statistically one-dimensional and fully developed, the equation governing the average velocity can be written as 
\begin{align}
  \nu\frac{\mathrm{d}\uave}{\mathrm{d}y} = \overline{u^\prime v^\prime} - \frac{\tau_{w}}{\rho}\frac{y}{h}
\end{align}
where 
\begin{align}
  \tau_{w} \equiv \rho\nu \left.\frac{\mathrm{d}\uave}{\mathrm{d}y}\right|_{y=-h}
\end{align}
is the wall shear stress.  A key observation here is that the average velocity is driven by the $u-v$ component of the Reynolds stress tensor.  Hence, if the goal is solely to predict the mean flow, then one only needs to worry about accurately predicting $b_{uv}$.  Higher order moments (such as energy) depend on other components of the Reynolds stress tensor.

It is natural to normalize the wall-normal distance $y$ and to work in viscous wall units, which are denoted by $y^{+}\equiv y/h_{\nu}$, 
where $h_{\nu} = \nu / u_{\tau}$ is the viscous lengthscale and $u_{\tau} = \sqrt{\tau_{w} / \rho}$ is called the friction velocity. The 
dimensionless number $\mathrm{Re}_{\tau} = u_{\tau}h/\nu$ is called the friction Reynolds number. Working with $y^{+}$ units is a convenient 
normalization in  channel flow as it naturally reveals important regions of the flow field.  In channel flow, the near-wall and the bulk 
regions exhibit distinctly different flow features, dissipation being mostly localized within the former.  The near-wall region occurs at 
about $y^{+} < 50$ and the bulk region occurs for $y^{+} > 50$.  Many simple eddy-viscosity models do not make any distinction between these 
regions and  ad-hoc ``wall-functions'' are often introduced into the models to account for the near-wall 
behavior~\cite{er1956turbulent, launder1972mathematical, pope2001turbulent}.  A machine learning model informed by direct numerical 
simulation should be intrinsically aware of the qualitatively different flow regions.

In the turbulent channel flow, the only non-zero component of the rate of strain tensor is $\mathrm{d}\uave/\mathrm{d}y$.  For ease of 
notation, we let 
\begin{align*}
   \alpha = \frac{k}{2\epsilon}\frac{\mathrm{d}\uave}{\mathrm{d}y}.
\end{align*}
Then, in terms of the general eddy viscosity model~\eqref{eq:nlev}, 
\begin{align}
    b_{uv} &= g^{(1)} \alpha - 2g^{(6)} \alpha^3 \label{eq:buv} \\
    b_{uu} &= -2g^{(2)}\alpha^2 + \frac{1}{3}\left( g^{(3)}-g^{(4)} \right)\alpha^2 - 2\left( g^{(7)}+g^{(8)} \right) \alpha^4 - \frac{2}{3}g^{(9)}\alpha^4 \label{eq:buu} \\
    b_{vv} &= 2g^{(2)}\alpha^2 + \frac{1}{3}\left( g^{(3)}-g^{(4)} \right)\alpha^2 + 2\left( g^{(7)}+g^{(8)} \right) \alpha^4 - \frac{2}{3}g^{(9)}\alpha^4 \label{eq:bvv} \\
    b_{ww} &= -\frac{2}{3}\left( g^{(3)}-g^{(4)} \right)\alpha^2 + \frac{4}{3} g^{(9)}\alpha^4 \label{eq:bww}.
\end{align}
All other components are identically zero. In comparison, the linear eddy viscosity model~\eqref{eq:levm_normalized} gives 
\begin{align}
    b_{uv} = -C_{\mu} \alpha
\end{align}
with all other components being zero.  From this, we observe that $C_{\mu}$ corresponds to $-g^{\lr{1}} + 2 g^{\lr{6}} \alpha^2 $. Note too, that in order to account for the diagonal components, higher order terms are needed.  Table~\ref{tab:activebasis} summarizes the active basis tensors of the GEVM for the channel flow.
\begin{table}[h]
    \centering
    \begin{tabular}{cc}
    \toprule
        Component of $\nanirs$ & Active basis tensors \\
    \midrule
        $b_{uv}$ &  $\mathbf{T}^{(1)}$, $\mathbf{T}^{(6)}$ \\
        $b_{uu}$ &  $\mathbf{T}^{(2)}$, $\mathbf{T}^{(3)}$, $\mathbf{T}^{(4)}$, $\mathbf{T}^{(7)}$, $\mathbf{T}^{(8)}$, $\mathbf{T}^{(9)}$ \\
        $b_{vv}$ &  $\mathbf{T}^{(2)}$, $\mathbf{T}^{(3)}$, $\mathbf{T}^{(4)}$, $\mathbf{T}^{(7)}$, $\mathbf{T}^{(8)}$, $\mathbf{T}^{(9)}$ \\
        $b_{ww}$ &  $\mathbf{T}^{(3)}$, $\mathbf{T}^{(4)}$, $\mathbf{T}^{(9)}$  \\
    \bottomrule
    \end{tabular}
    \caption{Active basis tensors for non-zero components of $\nanirs$ in channel flows. }
    \label{tab:activebasis}
\end{table}
As stated in section~\ref{sec:GEVM} the machine learning algorithm will be used to learn the coefficients in the tensor basis.  The next section describes a neural network machine learning approach for learning the coefficients.  Background and terminology on neural networks is provided before reviewing the tensor basis neural network proposed in~\cite{ling2016reynolds}.

\subsection{Neural networks}\label{sec:NN}
\begin{figure}[h!]
	\centering
	\includegraphics[width=0.75\textwidth]{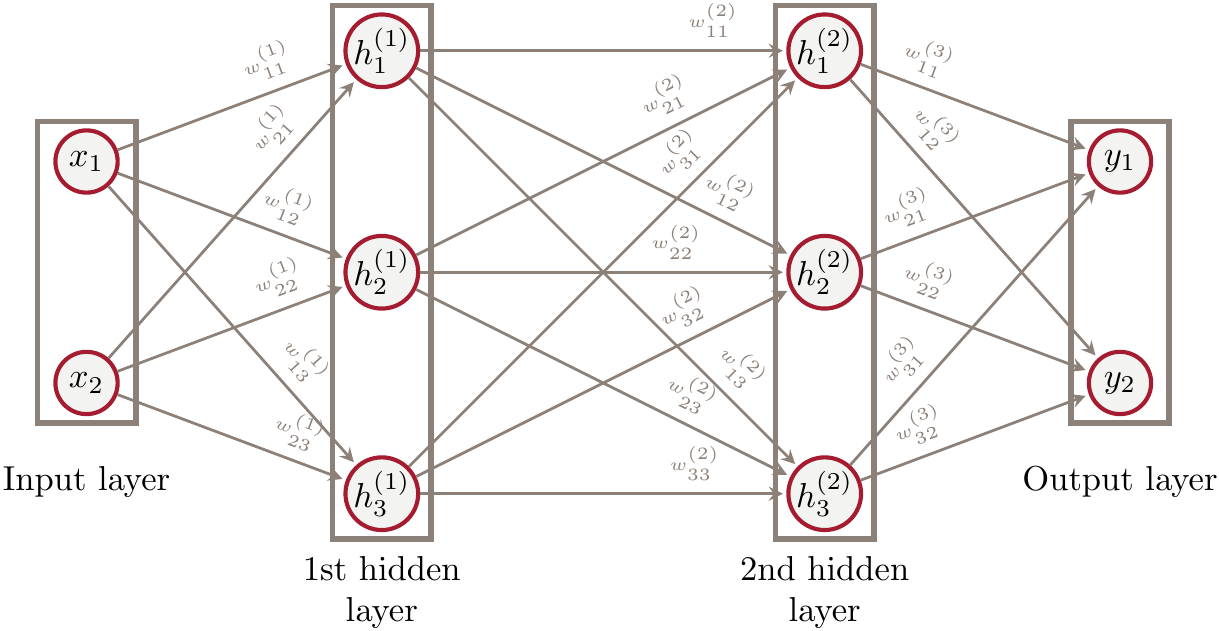}
    \caption{Diagram of a fully-connected feed-forward network with two hidden layers.}
    \label{fig:mlp_small}
\end{figure}

Neural networks are a class of machine learning algorithms that have found applications in a variety of fields, including computer 
vision~\cite{krizhevsky2012imagenet}, natural language processing~\cite{lecun2015deep}, and gaming~\cite{silver2017mastering}. Neural 
networks have shown to be particularly powerful in dealing with high-dimensional data and modeling nonlinear and complex relationships. 
Mathematically, a neural network defines a mapping $f: \mathbf{x} \mapsto \mathbf{y}$ where $\mathbf{x}$ is the input variable and 
$\mathbf{y}$ is the output variable. The function $f$ is defined as a composition of many different functions, which can be represented 
through a network structure.  As an example, figure~\ref{fig:mlp_small} depicts a basic fully-connected feed-forward network that defines a 
mapping $f: \mathbb{R}^2 \mapsto \mathbb{R}^2$.
The essential idea is outlined in the following enumeration.
\begin{enumerate}
    \item The input layer represents a 2-dimensional vector input $\mathbf{x} = [x_1, x_2]^\mathsf{T}$, with each node in the layer standing for each component of the vector.
    \item At the first hidden layer, the input $\mathbf{x}$ gets transformed into a 3-dimensional output $\mathbf{h^{\lr{1}}}$. This is done in two steps:
    \begin{enumerate}
        \item First, an affine transformation is performed at each node in the hidden layer:
            \begin{align*}
                z^{\lr{1}}_j = b^{\lr{1}}_j + \sum_{i=1}^{2} w^{\lr{1}}_{ij} x_i, \quad j = 1, 2, 3
            \end{align*}
        where $b^{\lr{1}}_j$ is the bias value for node $j$ and $w^{\lr{1}}_{ij}$ is the weight value associated with the arrow linking node $i$ in the input layer to node $j$ in the first hidden layer. 
        \item Second, a nonlinear transformation is performed according to a pre-specified (user-selected) activation function, $\phi$, 
            \begin{align*}
                h^{\lr{1}}_j = \phi\lr{ z^{\lr{1}}_j}
            \end{align*}
        An example of an activation function is the logistic function 
          \begin{align*}
              \phi\lr{z} = \dfrac{1}{1 + e^{-z}}
          \end{align*}
        Note that depending on the value of $z$ a node may output nothing if it is not activated (e.g. the limit as $z\to-\infty$).
    \end{enumerate}
    Altogether, in vector notation,
        \begin{align*}
            \mathbf{h^{\lr{1}}} = \phi\lr{\mathbf{W^{\lr{1}}} \mathbf{x} + \mathbf{b^{\lr{1}}}}
        \end{align*}
    where $\phi$ operates element-wise and the weight matrix $\mathbf{W^{\lr{1}}}$ and the bias vector $\mathbf{b^{\lr{1}}}$ are defined by 
        \begin{align*}
            \mathbf{W^{\lr{1}}}&=\begin{bmatrix}
                w^{\lr{1}}_{11} & w^{\lr{1}}_{12} &  w^{\lr{1}}_{13} \\
                w^{\lr{1}}_{21} & w^{\lr{1}}_{22} &  w^{\lr{1}}_{23} \\
            \end{bmatrix}^\mathsf{T}, \\
            \mathbf{b^{\lr{1}}}&=[b^{\lr{1}}_1, b^{\lr{1}}_2, b^{\lr{1}}_3]^\mathsf{T}.
        \end{align*}
    \item Similarly, the second hidden layer takes $\mathbf{h^{\lr{1}}}$ as input and produces a 3-dimensional output 
        \begin{align*}
            \mathbf{h^{\lr{2}}} = \phi\lr{\mathbf{W^{\lr{2}}} \mathbf{h^{\lr{1}}} + \mathbf{b^{\lr{2}}} }
        \end{align*}
    \item Finally, the output layer returns the 2-dimensional output of the network 
        \begin{align*}
            \mathbf{y} = \phi_{\textrm{out}} \lr{\mathbf{W^{\lr{3}}} \mathbf{h^{\lr{2}}} + \mathbf{b^{\lr{3}}} }
        \end{align*}
    The transformation $\phi_{\textrm{out}}$ is different from the nonlinear activation in the hidden layers. The choice of $\phi_{\textrm{out}}$ is guided by the output type and output distribution. For continuous outputs,  $\phi_{\textrm{out}}$ can simply be the identity in which case the output is a linear combination of the final hidden layer.  
\end{enumerate}
The network just described is an example of a fully-connected, feed-forward network.  It is fully-connected because every node in a hidden 
layer is connected with all the nodes in the previous and the following layers. It is feed-forward because the information flows in a 
forward direction from input to output; there is no feedback connection where the output of any layer is fed back into itself. A 
fully-connected, feed-forward network is the most basic type of neural network and is commonly referred to as a multilayer perceptron 
(MLP).  Interestingly, it has been mathematically proven that MLPs are universal function approximators \cite{hornik1989multilayer}.

\begin{figure}[h!]
	\centering
	\includegraphics[width=0.75\textwidth]{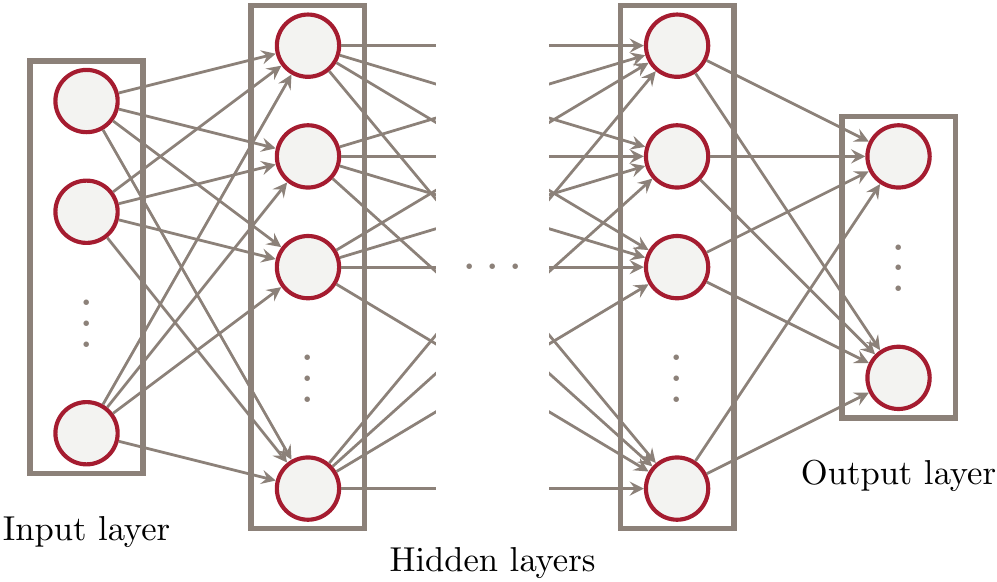}
    \caption{Diagram of a fully-connected feed-forward network.}
    \label{fig:mlp_generic}
\end{figure}

A generic MLP is shown in figure~\ref{fig:mlp_generic}.  The complexity of such a neural network increases with the number of hidden layers (depth of the network) and the number of nodes per hidden layer (width of the network). Networks with more than one hidden layer are called deep neural networks. 

\subsubsection{Training a neural network}\label{sec:training}
The neural network expresses a functional form $\hat{f}$ that is parameterized by a set of weights and biases, which are denoted by $W$. This functional form is an approximation to the true function $f$.  To find the best function approximation, one solves an optimization problem that minimizes the overall difference between $\hat{f}(\mathbf{x})$ and $f(\mathbf{x})$ for all $\mathbf{x}$ in the dataset to obtain the model parameters. The process of finding the best model parameters is called model training or learning.  Once the model is trained, its performance is assessed on the test dataset.  Training and test datasets are generated from the full dataset by splitting it into testing and training portions.  Often, the split is done with $20\%$ of the dataset used for testing and $80\%$ used for training. 

The overall difference between the true function and the approximation is quantified by a loss function.  Typically, the choice of loss function is dependent on the particular problem.  A general form of the total loss function is, 
    \begin{align*}
        \mathcal{L} \lr{W} = \frac{1}{N}\sum_{n=1}^N \mathcal{L}_n \lr{W}. 
    \end{align*}
where $N$ is the total number of data points used for training and $\mathcal{L}_n $ is the loss function defined for a single data point. A commonly used loss function is the mean squared error (MSE) loss 
    \begin{align*}
        \mathcal{L} \lr{W} = \frac{1}{N}\sum_{n=1}^N \left[ \lr{f\lr{\mathbf{x}_n} - \hat{f}\lr{\mathbf{x}_n}} \cdot \lr{f\lr{\mathbf{x}_n} - \hat{f}\lr{\mathbf{x}_n} } \right].
    \end{align*}
 
The stochastic gradient descent method and its variants are used to iteratively find parameters that minimize the loss function \cite{goodfellow2016deep}. 
In standard gradient descent, the model parameters $W$ are updated according to,
    \begin{align*}
        W^{k} &= W^{k-1} - \eta \nabla \mathcal{L} \lr{W^{k}} \\
                   &= W^{k-1} - \eta \lr{\frac{1}{N}\sum_{n=1}^N \nabla \mathcal{L}_n \lr{W^{k}}}
    \end{align*}
where $W_{k}$ are the model parameters at step $k$ and $\eta$ is the learning rate. This step repeats until convergence is achieved to within a user-specified tolerance.

Although neural networks have impressive approximation properties, training them requires the solution of a non-convex optimization problem.  The classical gradient descent algorithm has significant trouble in finding a global minimum and can often get stuck in a shallow local minimum.  The stochastic gradient descent algorithm provides a way of escaping from local minima in an effort to get closer to a global minimum.  In each iteration of the stochastic gradient decsent, the gradient $\nabla \mathcal{L} \lr{W}$ is approximated by the gradient at a single data point $\nabla \mathcal{L}_n \lr{W}$,
\begin{align*}
    W^{k} = W^{k-1} - \eta \nabla \mathcal{L}_n \lr{W^{k}}
\end{align*}
The algorithm sweeps through the training data until convergence to a local minimum is achieved. One full pass over the training data is called an epoch. Note that the training data is randomly shuffled at the beginning of each epoch.  This algorithm is stochastic in the sense that the estimated gradient using a random data point is noisy whereas the gradient calculated on the entire training data is exact. In practice, mini-batch stochastic grdient descent is employed, in which multiple data points are used in each iteration.  The batch size controls the number of random data points used per iteration.  For parameter initialization, in most cases the initial weights are randomly sampled from a uniform or normal distribution and the initial biases are set to 0. 

\subsubsection{Additional considerations}\label{sec:nn-comments}
The purview of neural networks is vast and growing.  In addition to the key aspects outlined above, there are a few additional considerations concerning neural networks that we outline presently.

Neural networks are prone to overfitting a dataset.  In this context, overfitting refers to the phenomenon whereby the network matches the 
training data very well but is unable to fit the test data; that is, the learned neural network does not generalize to the test set.  One 
approach to alleviate this issue is to add a regularization term to the loss function. For example, imposing an $L^2$ penalization term on 
the loss function constrains the magnitudes of the model parameters. For neural networks it is also popular to implement early-stopping in 
which a portion of the training data are held out as validation data and the validation error is monitored during training.  The training 
process terminates once the validation error begins to increase. 

Besides model parameters, the performance of a neural network changes with the external configuration of the network model and the training process.  The external configuration refers to the number of hidden layers, the number of nodes per layer, the activation function and the learning rate. These are called the hyperparameters of a model. The search for the best values of hyperparameters is called hyperparameter tuning. A grid search can be performed to search combinations of values on a grid of parameters in the hyperparameter space. A separate validation set that is different from the test set is used for model evaluation during the tuning process.  Alternatively, a Bayesian optimization \cite{snoek2012practical} of the hyperparameters may also be performed.  Table~\ref{tab:keyterms} summarizes the key terminology just introduced. 
    \begin{table}[h!]
        \centering
        \begin{tabular}{|M{0.3\linewidth}|M{0.65\linewidth}|}
            \hline 
                \textbf{Term} & \textbf{Explanation}  \\ \hline
            fully-connected, feed-forward network & basic type of neural network \\ \hline
            layer & a vector-valued variable serving as input, output, or intermediate output (in which case termed ``hidden layer'') in a neural network \\ \hline
            node & individual element of a vector represented by layer \\ \hline
            activation function  &  nonlinear function performed on nodes in hidden layers \\ \hline 
            model parameters & weights and biases tuned during the training process  \\ \hline 
            loss function  &  a scalar-valued function to be minimized during the training process \\ \hline
            stochastic gradient descent & a commonly used optimization algorithm for training neural networks \\ \hline
            learning rate &  step size of the iterative gradient-based optimization algorithm \\ \hline
            epoch & a full pass through all training data in stochastic algorithms \\ \hline
            batch size & number of data points used to estimate gradients in one iteration of the stochastic algorithm \\ \hline
            model hyperparameters & external configuration of a network and the training process, such as the number of hidden layers, number of nodes per layer, activation function, learning rate, etc. \\ \hline
            train, validation, test sets & the whole dataset is split into train, validation and test sets for training, tuning and evaluating a model \\ \hline
            $L^2$ penalization & a regularization term added to the loss function in order to prevent overfitting \\ \hline
            early-stopping & a regularization technique that controls the training time in order to prevent overfitting \\ \hline 
        \end{tabular}
        \caption{Key terminology of neural networks.}
        \label{tab:keyterms}
    \end{table}

\subsection{The tensor basis neural network} \label{sec:bg-tbnn}
\begin{figure}[h!]
	\centering
	\includegraphics[width=0.6\textwidth]{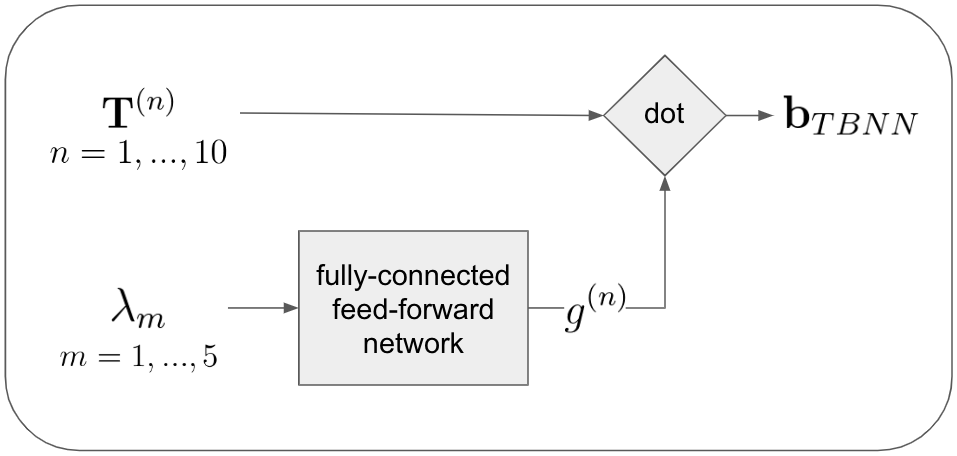}
    \caption{Diagram of the tensor basis neural network.}
    \label{fig:tbnn}
\end{figure}

With the terminology introduced in the last section, we are now ready to introduce the deep neural network proposed in \cite{ling2016reynolds}.  A schematic of the tensor basis neural network (TBNN) is provided in figure~\ref{fig:tbnn}. The TBNN consists of two input layers.  The first input layer is given by the scalar invariants~\eqref{eq:invariants} and the second input layer is the actual tensor basis components~\eqref{eq:basistensors}.  

The inputs to the neural network are intended to be derived from a RANS flow field (e.g. $\vave \lr{\mathbf{x}, t}$). The preprocessing 
procedure involves:
\begin{enumerate}
    \item Calculate the normalized mean rate of strain and rotation $\Shat\lr{\mathbf{x}, t}$ and $\Rhat\lr{\mathbf{x}, t}$ from $\vave 
          \lr{\mathbf{x}, t}$, $k \lr{\mathbf{x}, t}$, $\epsilon \lr{\mathbf{x}, t}$ following \eqref{eq:nstrainrotation};
    \item Calculate and the five scalar invariants $\lambda_m\lr{\mathbf{x}, t}$ and the ten basis tensors 
          $\mathbf{T}^{(n)}\lr{\mathbf{x},t}$ from $\Shat\lr{\mathbf{x}, t}$ and $\Rhat\lr{\mathbf{x}, t}$ following \eqref{eq:invariants} 
          and \eqref{eq:basistensors}.
\end{enumerate}
The five scalar invariants are fed through a fully-connected feed-forward network, the output of which is the tensor basis coefficients 
$g^{\lr{n}}, n=1,2,...,10$.  These are then combined with the ten basis tensors to form the normalized anisotropy tensor 
$\nanirs\lr{\mathbf{x}, t}$, according to~\eqref{eq:nlev}. The true values of $\nanirs\lr{\mathbf{x}, t}$ are provided by DNS data of the 
same flow. Specific details of the architecture can be found in the original reference~\cite{ling2016reynolds}.

\begin{figure}[h!]
	\centering
	\includegraphics[width=0.4\textwidth]{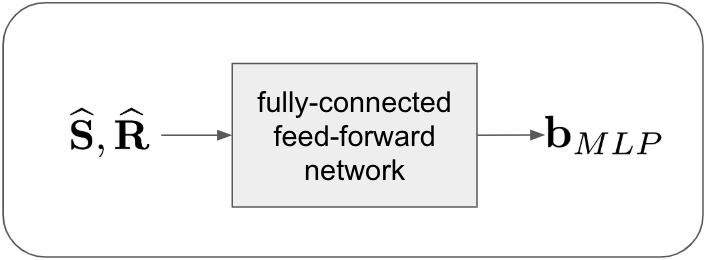}
    \caption{Diagram of the MLP used to predict the normalized anisotropy tensor. }
    \label{fig:mlp}
\end{figure}

In~\cite{ling2016reynolds}, the authors trained, validated and tested the TBNN on a total of nine flows: six for training (duct flow, channel flow, a perpendicular jet in cross-flow, an inclined jet in cross-flow, flow around a square cylinder, flow through a converging-diverging channel), one for validation (a wall-mounted cube in cross-flow), and two for test (duct flow, flow over a wavy wall).  They compared the Reynolds stress anisotropy predictions of the TBNN with those of the default linear eddy viscosity model (LEVM), a quadratic eddy viscosity model (QEVM) \cite{craft1996development} and a fully-connected feed-forward network (MLP). As illustrated in figure~\ref{fig:mlp}, the MLP predicts $\nanirs\lr{\mathbf{x}, t}$ from the nine distinct components of $\Shat\lr{\mathbf{x}, t}$ and $\Rhat\lr{\mathbf{x}, t}$.  
The authors showed that the TBNN provided the best results when compared to the LEVM, QEVM, and MLP. They also explored whether the improved anisotropy predictions would translate to improved mean velocity predictions, by inserting the TBNN predicted Reynolds stress anisotropy values into an in-house RANS solver for the two test cases.  This evaluation showed that the TBNN was capable of capturing key flow features that the LEVM and QEVM both failed to predict, including a separation bubble.

\section{Methodology} \label{sec:methods}
In the present work, we analyze turbulent channel flow, which was one of the flows considered in~\cite{ling2016reynolds}.  Rather, than 
extend the results in~\cite{ling2016reynolds}, our focus is to assess how the TBNN learns the turbulence physics encoded within the tensor basis representation.

\subsection{Dataset}
We trained and evaluated the TBNN and the MLP using a channel flow DNS~\cite{lee2015direct} at a friction Reynolds number $\Rey_\tau = 1000$.  The raw data were the mean velocity gradients $\nabla\uave\lr{y^{+}}$, the turbulent kinetic energy $k\lr{y^{+}}$, the turbulent dissipation $\epsilon\lr{y^{+}}$ and the Reynolds stresses $\mathcal{R}\lr{y^{+}}$ derived from the DNS.  

Using $\mathcal{R}\lr{y^{+}}$ and $k\lr{y^{+}}$, we computed the normalized anisotropy tensor $\nanirs\lr{y^{+}}$ according to \eqref{eq:anisotropy} and \eqref{eq:normalized_anisotropy}, which were then used as truth labels.  As stated in section~\ref{sec:bg-tbnn}, the inputs to the neural network should be RANS data. Given that we only had the DNS data, we generated synthetic RANS data by smoothing the DNS fields $\nabla\uave\lr{y^{+}}$, $k\lr{y^{+}}$, $\epsilon\lr{y^{+}}$ with a moving average filter of width $3$. We then calculated the inputs to the TBNN and the MLP following the preprocessing procedure described in section~\ref{sec:bg-tbnn}.  For the TBNN, the inputs were the five scalar invariants $\lambda_m\lr{y^{+}}$ and the ten basis tensors $\mathbf{T}^{(n)}\lr{y^{+}}$. For the MLP, the inputs were the nine distinct components of the normalized mean rate of strain and rotation $\Shat\lr{y^{+}}$ and $\Rhat\lr{y^{+}}$.  The DNS provided $256$ data points over the wall-normal direction. Therefore, we had $256$ data points in total, which we split into $80\%$ training set and $20\%$ test set.  To summarize, tables~\ref{tab:datashapes_tbnn} and~\ref{tab:datashapes_mlp} present the shapes of the input and output data for the TBNN and the MLP. 
\begin{table}[h!]
    \centering
    \begin{tabular}{|c|c|c|c|}
        \hline
              & Inputs 1 ($\lambda_m\lr{y^{+}}, m=1,2,...,5$) & Inputs 2 ($\mathbf{T}^{(n)}\lr{y^{+}}, n=1,2,...,10$) & outputs ($\nanirs\lr{y^{+}}$)\\
        \hline
        Train & (204, 5) & (204, 10, 9) & (204, 9) \\
        Test  & (52, 5) & (52, 10, 9) & (52, 9) \\
        \hline
    \end{tabular}
    \caption{Input and output data shapes for the TBNN.}
    \label{tab:datashapes_tbnn}
\end{table}
\begin{table}[h!]
    \centering
    \begin{tabular}{|c|c|c|}
        \hline
              & Inputs (6 from $\Shat\lr{y^{+}}$, 3 from $\Rhat\lr{y^{+}}$) & outputs ($\nanirs\lr{y^{+}}$)\\
        \hline
        Train & (204, 9) & (204, 9) \\
        Test  & (52, 9) & (52, 9) \\
        \hline
    \end{tabular}
    \caption{Input and output data shapes for the MLP.}
    \label{tab:datashapes_mlp}
\end{table}

\subsection{Models} 
We implemented the TBNN and the MLP\footnote{\url{https://github.com/fr0420/machine-learning-turbulence}} in Pytorch \cite{paszke2017automatic} using the core package\footnote{\url{https://github.com/tbnn/tbnn}} originally developed in Theano \cite{al2016theano} by~\cite{ling2016reynolds}.  Our reimplementation was partly motivated by the desire to take advantage of a machine learning library that is under active development.  We compared the performance of both models with two traditional turbulence models (LEVM and QEVM). 


\subsection{Training}
The predicted output of the neural network is denoted by $\widehat{\nanirs}$ and the true value from the DNS is denoted by $\nanirs$. To accurately predict $\nanirs$, for the TBNN we defined a loss function 
\begin{align}
    \mathcal{L} = \frac{1}{6 N}\sum_{n=1}^{N} \sum_{ij\in \mathcal{I}}  l \lr{ b_{ij,n}, \widehat{b}_{ij,n}}
    \label{eq:b_loss_tbnn}
\end{align}
where $\mathcal{I} = \{uu,uv,uw,vv,vw,ww\}$, and $l\lr{\cdot, \cdot}$ is a function that defines the difference between the predicted and true values of a single component $b_{ij}$ for a single data point.  For instance, given two scalars $a$ and $b$, choosing $l \lr{a, b} = \lr{a-b}^2$ defines the mean-squared-error loss. Using the MSE loss is theoretically supported if the distribution of outputs is Gaussian. In the present work, this assumption is no longer valid.  Nevertheless, in the absence of theory and in the interest of simplicity, we follow this convention. 

\begin{figure}[h!]
    \centering
    \includegraphics[width=0.8\textwidth]{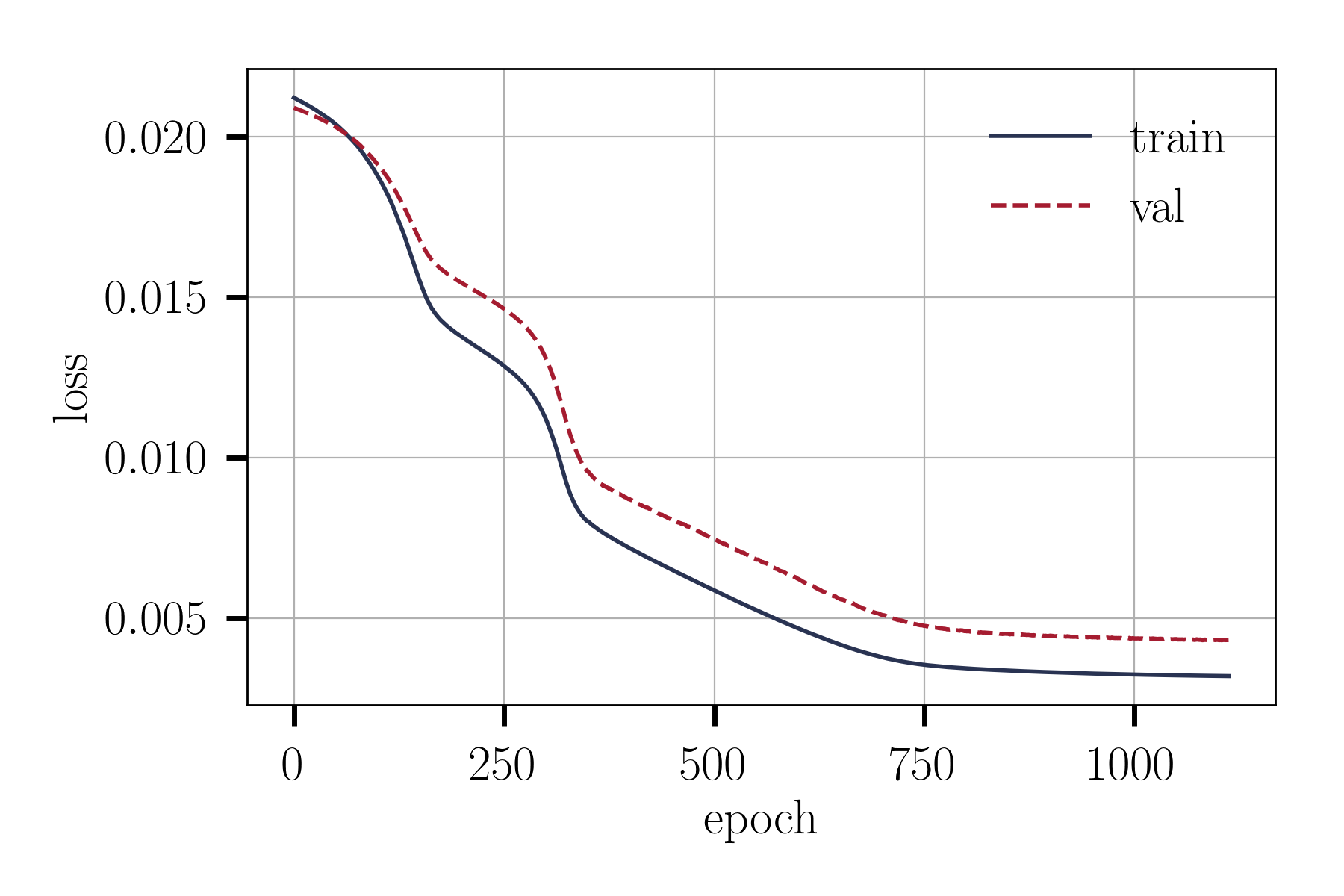}
    \caption{The training and validation loss as a function of training epochs. The curves shown here are most representative to averages of ten repeated runs with different randomly selected validation points. }
    \label{fig:train_val_curves}
\end{figure}

For the MLP, the loss function is defined on all nine components of $\nanirs$ since we did not enforce symmetry on the the predicted $\nanirs$.  For this case, the loss function is
\begin{align}
    \mathcal{L} = \frac{1}{9 N}\sum_{n=1}^{N} \sum_{i\in \{ u,v,w\}} \sum_{j\in \{ u,v,w\}}  l \lr{ b_{ij,n}, \widehat{b}_{ij,n} }
    \label{eq:b_loss_mlp}
\end{align}
As discussed in section \ref{sec:channelflow}, to predict the mean flow of a channel flow, only the $u-v$ component of $\nanirs$ is needed. Therefore, we defined an additional loss function for the TBNN that corresponds to only accurately predicting $b_{uv}$: 
\begin{align}
    \mathcal{L}_{uv} = \frac{1}{N}\sum_{n=1}^{N} l \lr{b_{uv, n}, \widehat{b}_{uv, n}}
    \label{eq:buv_loss}
\end{align}
Our main experiments were concerned with accurately predicting the entire tensor $\nanirs$. Hence, \eqref{eq:b_loss_tbnn} for TBNN and \eqref{eq:b_loss_mlp} for MLP were used. We used early-stopping as a regularization during training. Figure~\ref{fig:train_val_curves} shows an example of the training and validation loss as a function of epochs during training the TBNN. 



\subsection{Hyperparameter tuning} \label{sec:hyperparam}
For each neural network (TBNN and MLP), we examined ten hyperparameters: 
\begin{itemize}
    \item number of hidden layers
    \item number of nodes per hidden layer
    \item activation function
    \item loss function
    \item optimization algorithm 
    \item learning rate
    \item batch size
    \item $L^2$ penalization coefficient 
    \item weight initialization function 
    \item patience for early-stopping (number of epochs with no improvement in validation loss after which training will be stopped) 
\end{itemize}
Our objective is to find the hyperparameters that give the lowest loss on the validation set. In addition, the optimal model must not be too sensitive to the random state of the initial weights. 

We first explored the effect of each hyperparameter by varying one hyperparameter at a time while keeping the others fixed. The default 
hyperparameters were based on those used in \cite{ling2016reynolds}.  To ensure the model is not too sensitive to the random initial 
weights, for each choice of hyperparameters, we repeated the experiment 20 times using different random seeds for weight initialization, and 
then calculated the mean and variance of the loss scores on the validation set. Among the hyperparameters that produced a variance below a 
threshold of $0.01$, we selected the one yielding the lowest mean validation loss.  After that, we did a finer search 
for a subset of the hyperparameters. A grid search was performed to optimize the values of the number of hidden layers, the number of nodes 
per hidden layer, and the learning rate. With the optimal model, the validation loss improved from 0.0016 to 0.0009, corresponding to the 
$R^2$ score increasing from 0.40 to 0.65.  Table~\ref{tab:hyperparams} shows the optimal hyperparameters for the TBNN model trained to 
optimize the loss function \eqref{eq:b_loss_tbnn} 
\begin{table}[h!]
        \centering
        \begin{tabular}{cc}
            \toprule
                Name & Value  \\
            \midrule
            number of hidden layers  & 25 \\ 
            number of nodes per hidden layer & 100 \\
            activation function  &  Swish \cite{ramachandran2018searching} \\
            loss function  &  MSE (Mean squared error) \\
            optimization algorithm & Adam \\
            learning rate & $2.5\times 10^{-6}$\\ 
            batch size  &  10 \\
            $L^2$ penalization coefficient  &  0 \\
            weight initialization function  &  Xavier normal \\
            patience for early-stopping & 30  \\
            \bottomrule
        \end{tabular}
        \caption{Hyperparameter setting.}
        \label{tab:hyperparams}
    \end{table}


\section{Results} \label{sec:results}
\subsection{Comparison of $\nanirs$ profiles} \label{sec:profiles}
Table~\ref{tab:model_performance_r2} shows the $R^2$ values for the optimal TBNN and MLP models as well as the traditional LEVM and QEVM models.  Each column represents the performance on one component in the stress tensor except for the first column, which represents the performance on the entire stress tensor.  Inspection of the values in the table indicate that the TBNN and MLP models generally outperform the LEVM and QEVM models. 

    \begin{table}[h!]
        \centering
        \begin{tabular}{cccccc}
            \toprule
                 & $R^2_{\nanirs}$ & $R^2_{uu}$ & $R^2_{uv}$ & $R^2_{vv}$ & $R^2_{ww}$  \\
            \midrule
            TBNN     & 0.6067	& 0.6010	& 0.9390	& 0.4985	& 0.7334 \\ 
            MLP      & 0.7825	& 0.8400	& 0.9348	& 0.5600	& 0.9404 \\
            LEVM     & -7.5310  & -6.2462	& -25.6563	& -4.9440	& -7.0633 \\
            QEVM     & -40.8737 & -45.9671	& -25.6563	& -27.7132	& -79.9124  \\
            \bottomrule
        \end{tabular}
        \caption{$R^2$ of the TBNN, MLP, LEVM, and QEVM predictions. }
        \label{tab:model_performance_r2}
    \end{table}

\begin{figure}[h!]
  \centering 
  \includegraphics[width=0.8\textwidth]{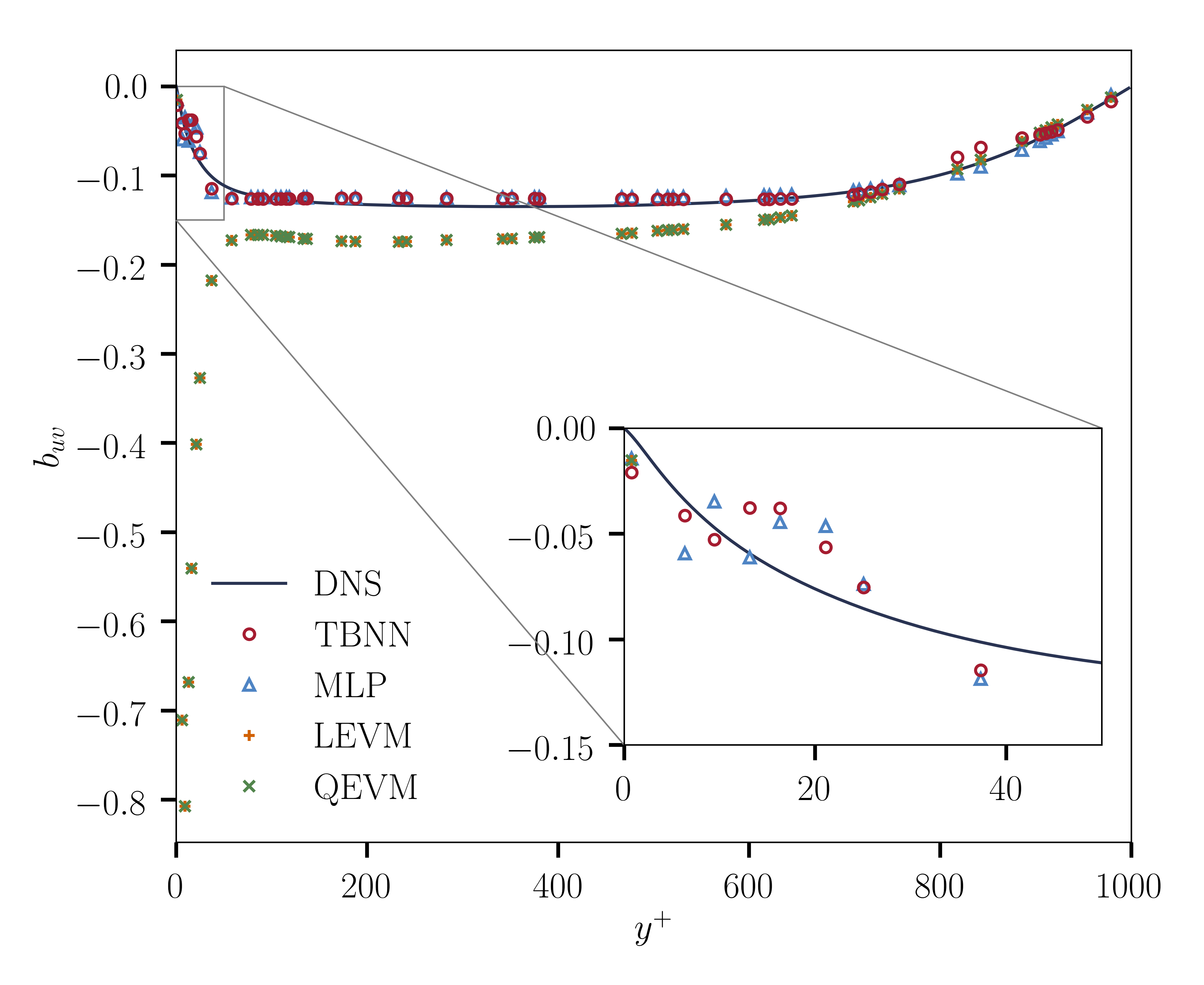}
  \caption{$u-v$ component of $\nanirs$.  The TBNN model is in good agreement with the DNS data. }
  \label{fig:buv}
\end{figure}

To gain a more qualitative picture of the model performance, we plot profiles of $\nanirs$.  Figure~\ref{fig:buv} reports $b_{uv}$ from the DNS data, the LEVM, the QEVM, the MLP, and the TBNN. 
For this flow field, the LEVM and the QEVM yield identical expressions for $b_{uv}$ and therefore provide the same erroneous prediction.  The LEVM and QEVM have the correct trend near the middle of the channel, but have the completely wrong behavior in the near-wall region.  The TBNN and MLP models perform the best, matching the DNS even in the near-wall region.  However, any model trained with the MLP will not automatically preserve the invariance properties, leaving predictive capabilities on other flow fields under question.  Additionally, even though it is easy to enforce symmetry manually, the MLP does not automatically guarantee symmetry of the Reynolds stress tensor; nor does it preserve the known invariance. In other words, while the TBNN is ``physics-aware", at least versus the basic symmetries, MLP is not.  This was shown to have important implications when applying the TBNN and MLP models to different flow fields~\cite{ling2016reynolds}.

\begin{figure}[h!]
        \centering
        \begin{subfigure}[b]{0.6\textwidth}
            \includegraphics[width=\textwidth]{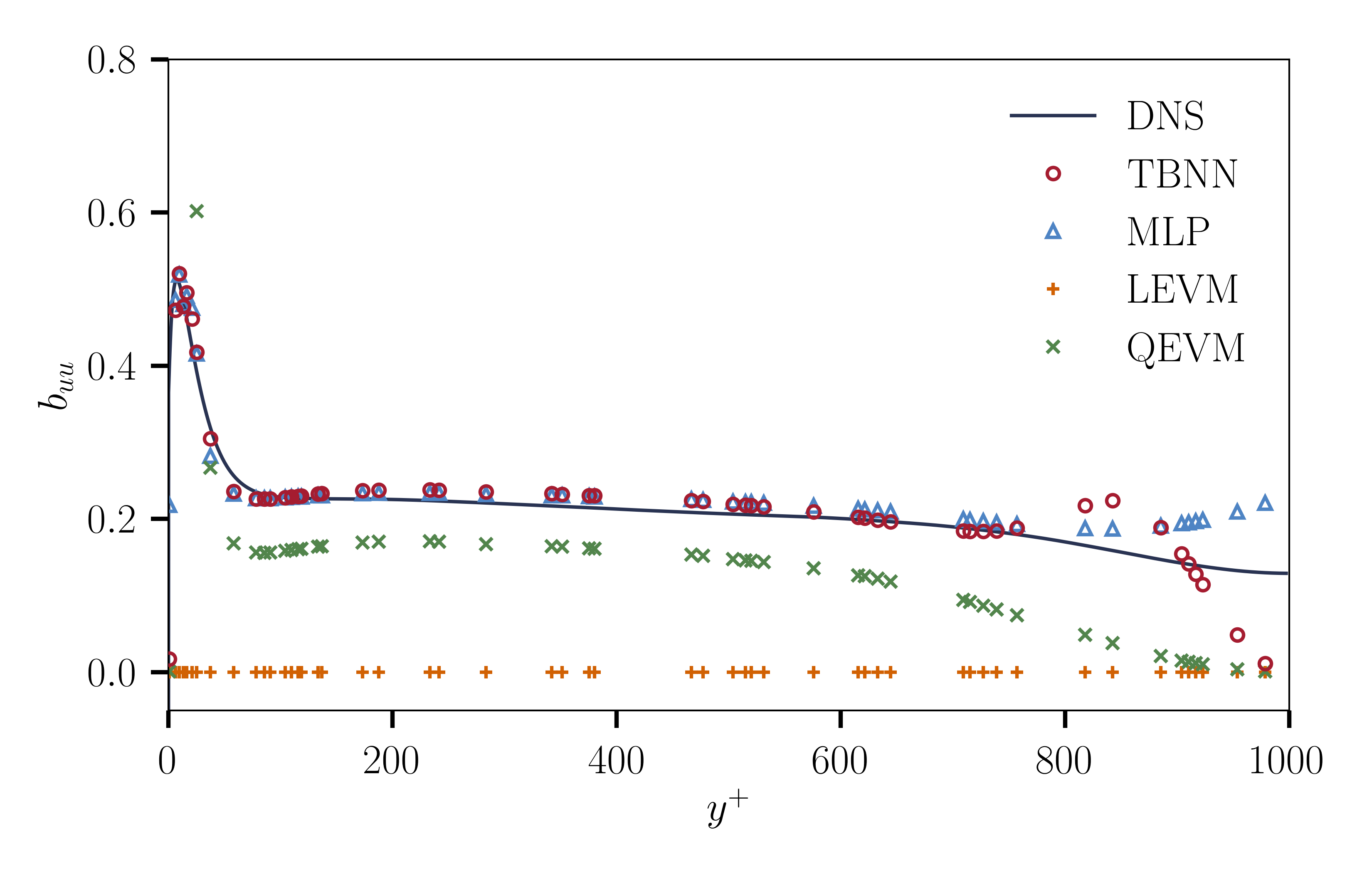}
            \caption{$b_{uu}$}
            \label{fig:buu}
        \end{subfigure}

        \begin{subfigure}[b]{0.6\textwidth}
            \includegraphics[width=\textwidth]{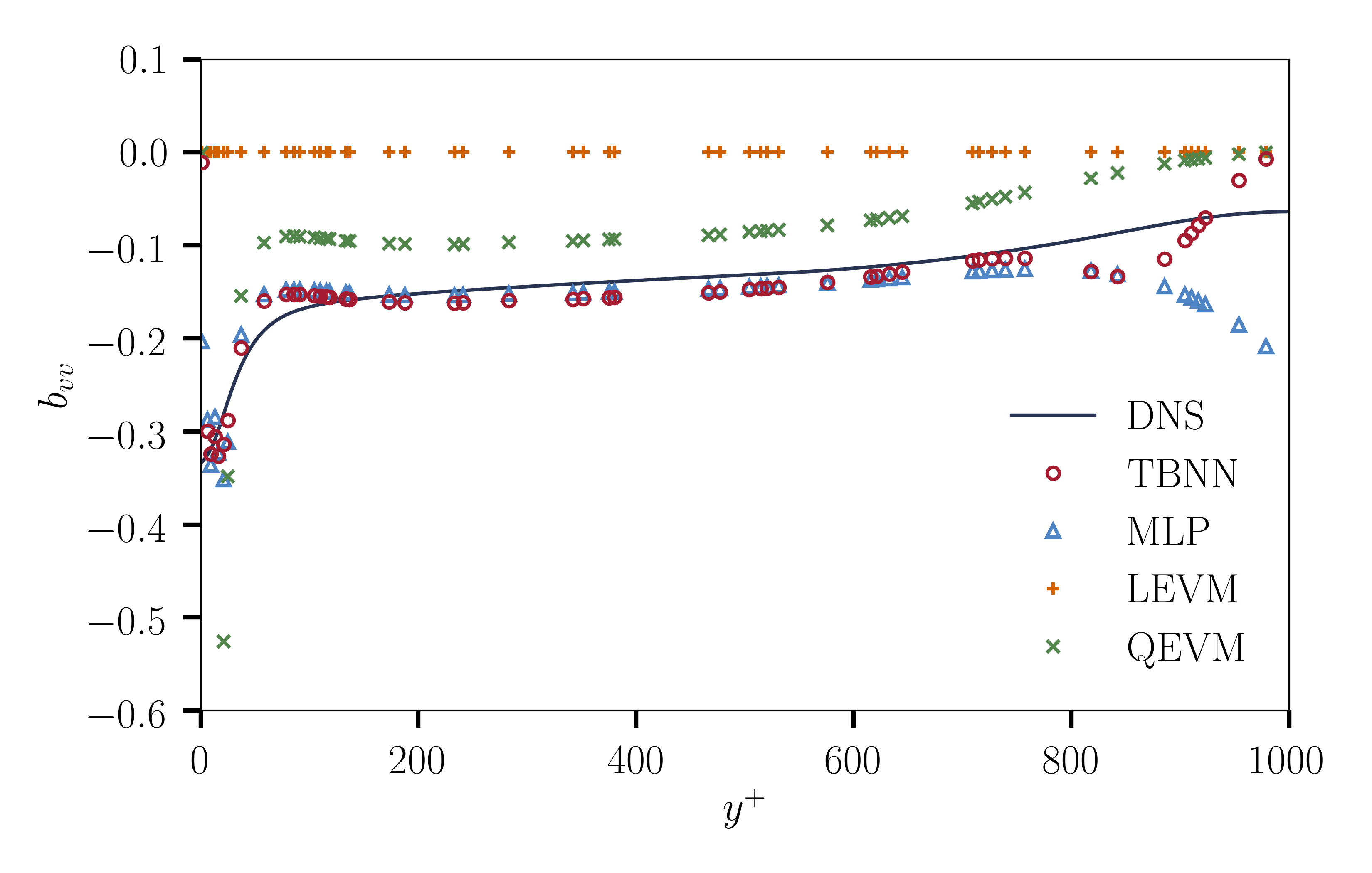}
            \caption{$b_{vv}$}
            \label{fig:bvv}
        \end{subfigure}

        \begin{subfigure}[b]{0.6\textwidth}
            \includegraphics[width=\textwidth]{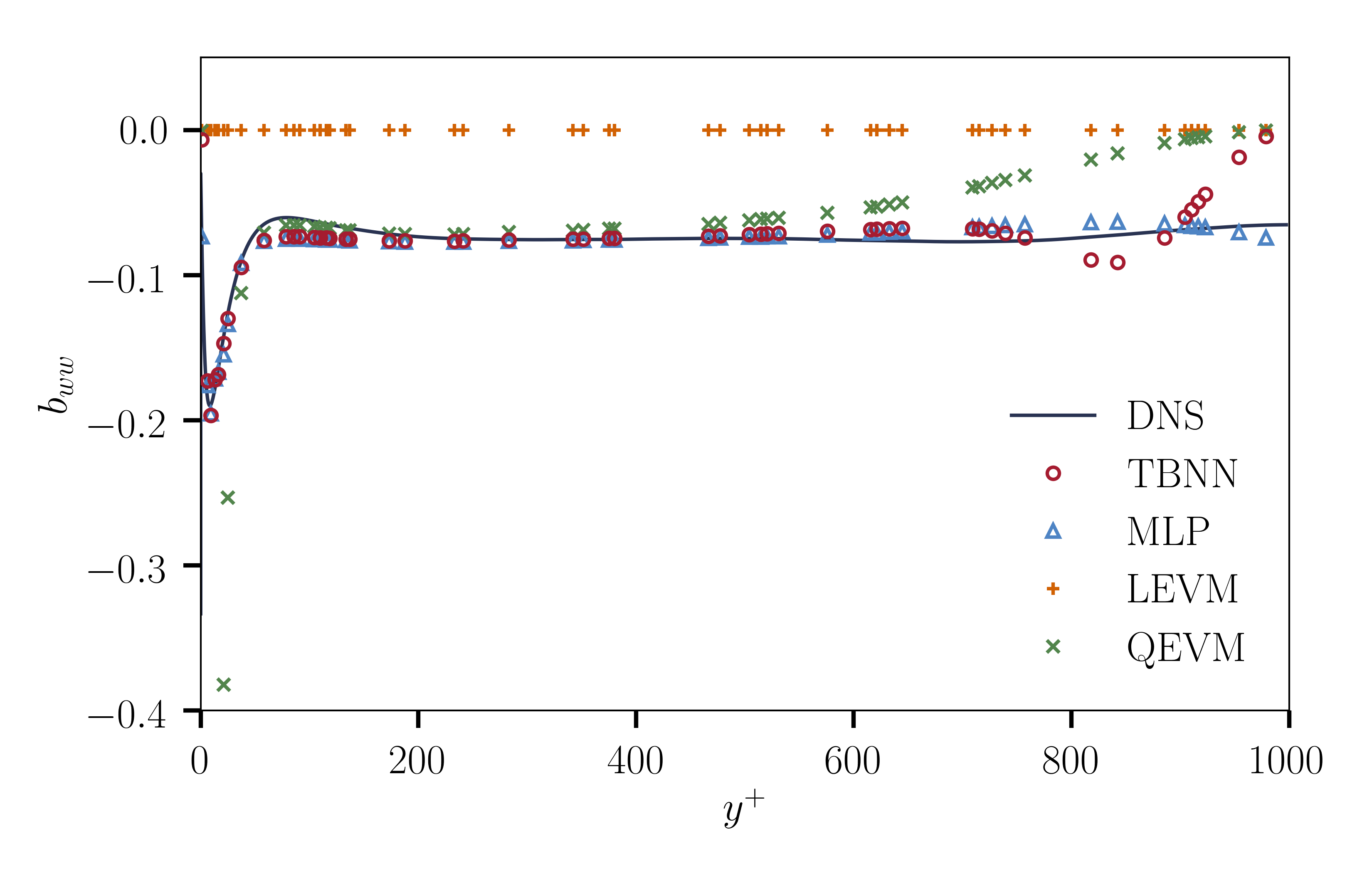}
            \caption{$b_{ww}$}
            \label{fig:bww}
        \end{subfigure}
        \caption{Comparison of the normal components of $\nanirs$.  The TBNN performs well, but has a deficiency at the center of channel due to the form of the GEVM~\ref{eq:nlev}.  The MLP responds to the DNS data near the center of the channel. The LEVM provides no predictions. The QEVM predicts the bulk region relatively well but fails completely in the near-wall region. }
        \label{fig:bnormal}
    \end{figure}

Figure~\ref{fig:bnormal} compares the normal components of $\nanirs$ from the DNS data to predictions from the various models considered in this work.
The TBNN model still performs well, although near the center of the channel the model is identically zero.  This is not necessarily a TBNN failure, but rather an inherent limitation of the GEVM~\eqref{eq:nlev} representation.  Indeed, by construction any model with an algebraic dependence on $\mathbf{S}$ and $\mathbf{R}$ only, is identically zero at the center of the channel.  DNS data, however, clearly show that for the channel flow this is not the case.  Hence, the GEVM representation is incomplete and the network built upon such a formulation will inevitably inherit this deficiency.  It may be instructive to inspect how the learning process ``tries" to cope with the problem, if at all.  On the other hand, the MLP model has no such constraint and, as shown in Figure~\ref{fig:bnormal}, it is able to respond to the DNS data near the center of the channel.  The LEVM, by definition, assumes these components are zero and therefore provides no prediction whatsoever. The QEVM, which contains nonlinear terms involving $\mathbf{S}$ and $\mathbf{R}$ in the formulation, does better than the LEVM in that it captures the correct trend in the bulk region of the channel. However, it fails completely in the near-wall region.  

\begin{figure}[h!]
        \centering
        \includegraphics[width=0.8\textwidth]{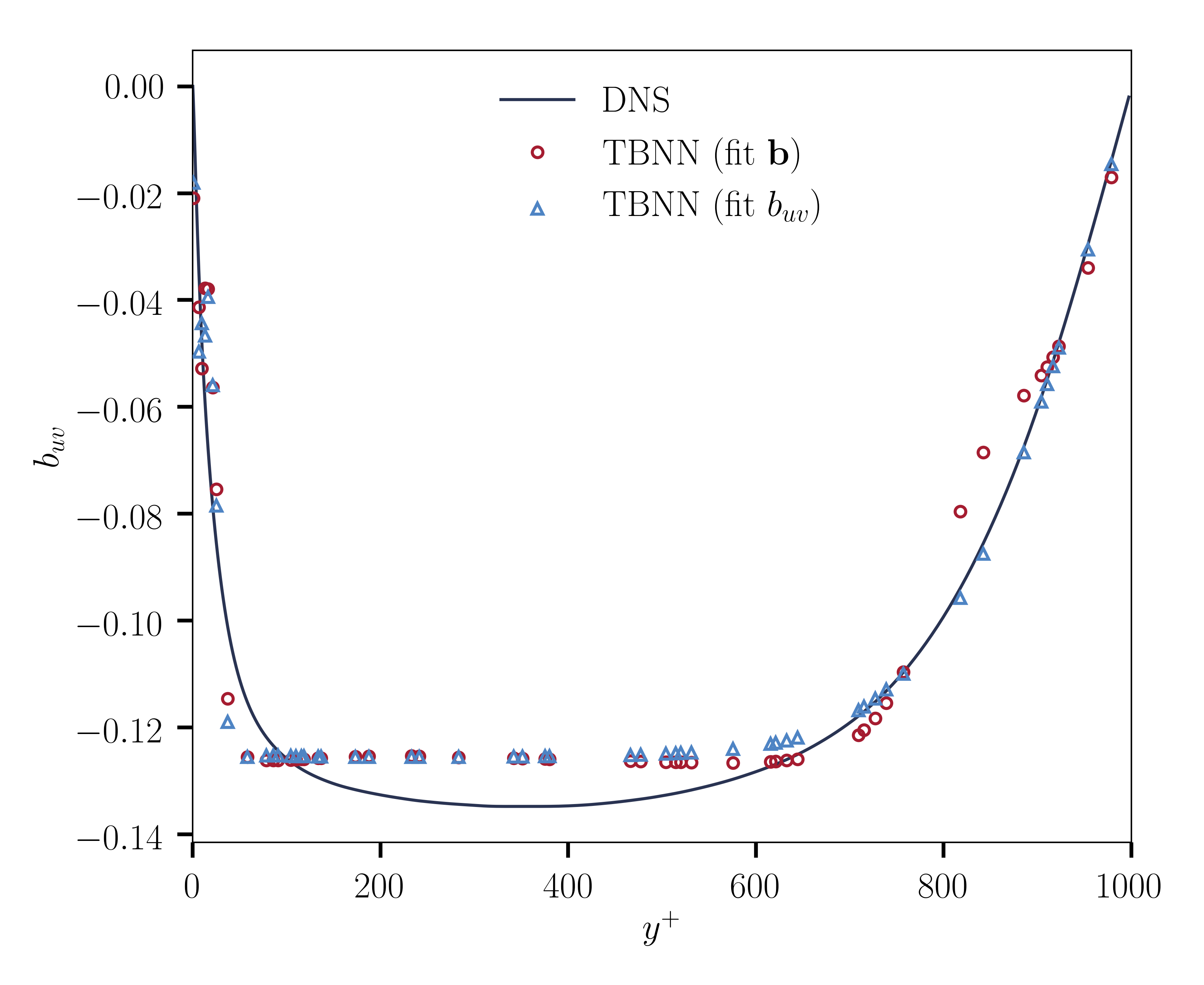}
        \caption{Predictions of $b_{uv}$ from two TBNN models: one being fit on the entire tensor $\nanirs$, the other one being fit on the $u-v$ component of the tensor $b_{uv}$. The results are better when the TBNN is trained on the $u-v$ component.}
        \label{fig:buv_TBNN}
    \end{figure}

The TBNN model presented above was trained on the entire stress tensor $\nanirs$. For comparison, we also trained a TBNN model on solely the $u-v$ component of the stress tensor $b_{uv}$. Table~\ref{tab:model_performance_r2_fitbuv} shows the performance of the two TBNN models on the $u-v$ component. 
    \begin{table}[h!]
        \centering
        \begin{tabular}{cccccc}
            \toprule
                 & $R^2_{uv}$  \\
            \midrule
            TBNN (fit $\nanirs$)    & 0.9390	 \\ 
            TBNN (fit $b_{uv}$)     & 0.9485	 \\ 
            \bottomrule
        \end{tabular}
        \caption{$R^2$ of $b_{uv}$ predictions from two TBNN models: one being fit on the entire tensor $\nanirs$, the other one being fit on the $u-v$ component of the tensor $b_{uv}$. }
        \label{tab:model_performance_r2_fitbuv}
    \end{table}
The $R^2$ values indicate that training on the $u-v$ component of the tensor can achieve better performance than training on the entire tensor.  This is also evident from figure~\ref{fig:buv_TBNN}, which shows the $b_{uv}$ profiles predicted by the TBNN models trained on the entire tensor and only on the $u-v$ component. Noticeably, the $b_{uv}$ predictions by the TBNN trained on the entire tensor are worse than the predictions by the TBNN trained only on the $u-v$ component at around $y^+=800$. This coincides with where the predictions of the normal components begin to fail (recall figure~\ref{fig:bnormal}) as a response to the inherent limitation of the GEVM~\eqref{eq:nlev} representation. In other words, in order to fit the whole tensor, the TBNN trained on the entire tensor may end up compromising the accuracy of the $b_{uv}$ predictions. 

\subsection{Connection with LEVM}
\begin{figure}[h!]
    \centering
    \begin{subfigure}[b]{0.7\textwidth}
        \includegraphics[width=\textwidth]{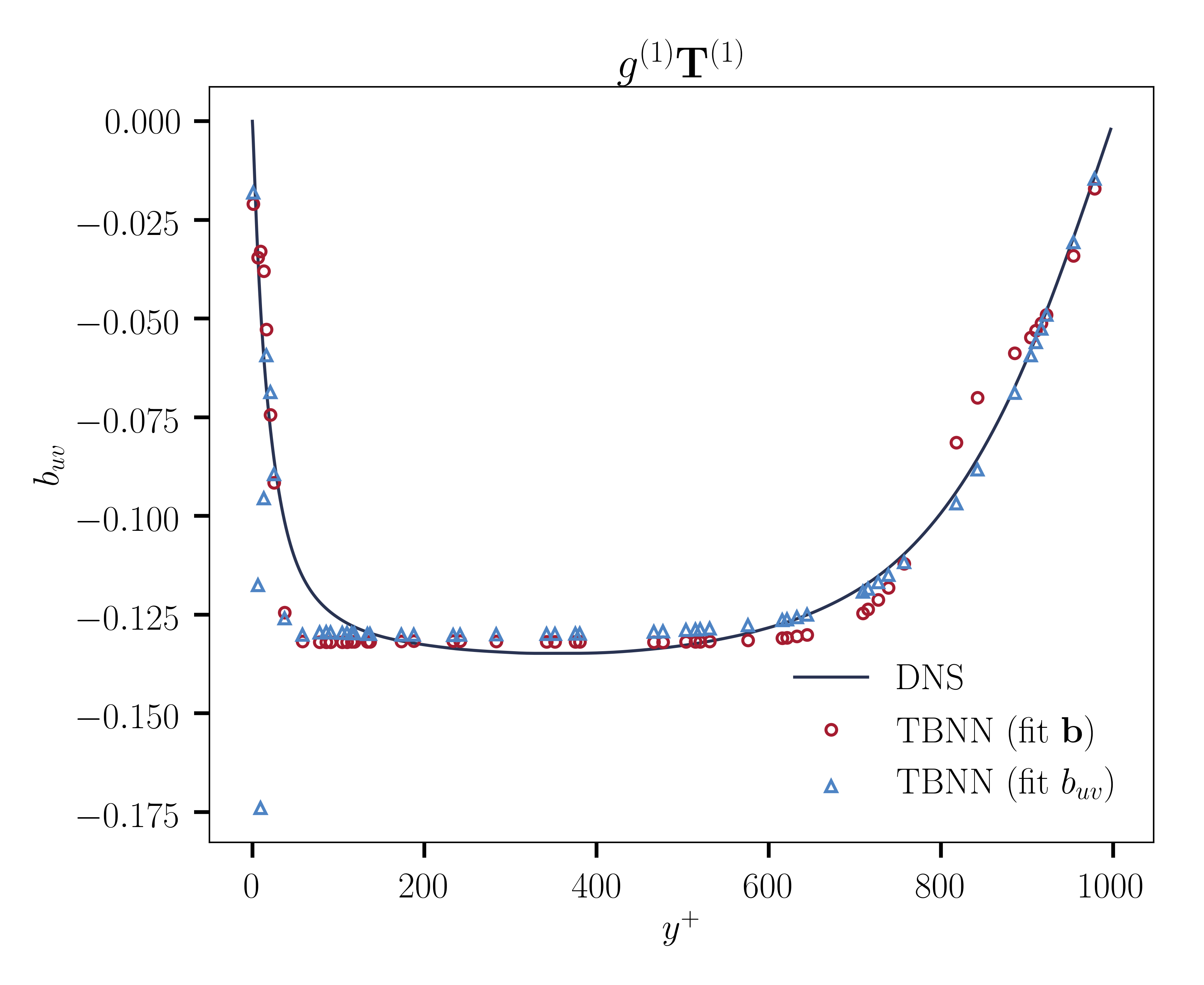}
        \caption{Predictions of term $g^{\lr{1}} \mathbf{T}^{\lr{1}}$ in the $u-v$ component.}
        \label{fig:buv_TBNN_g1T1}
    \end{subfigure}
    
    \begin{subfigure}[b]{0.7\textwidth}
        \includegraphics[width=\textwidth]{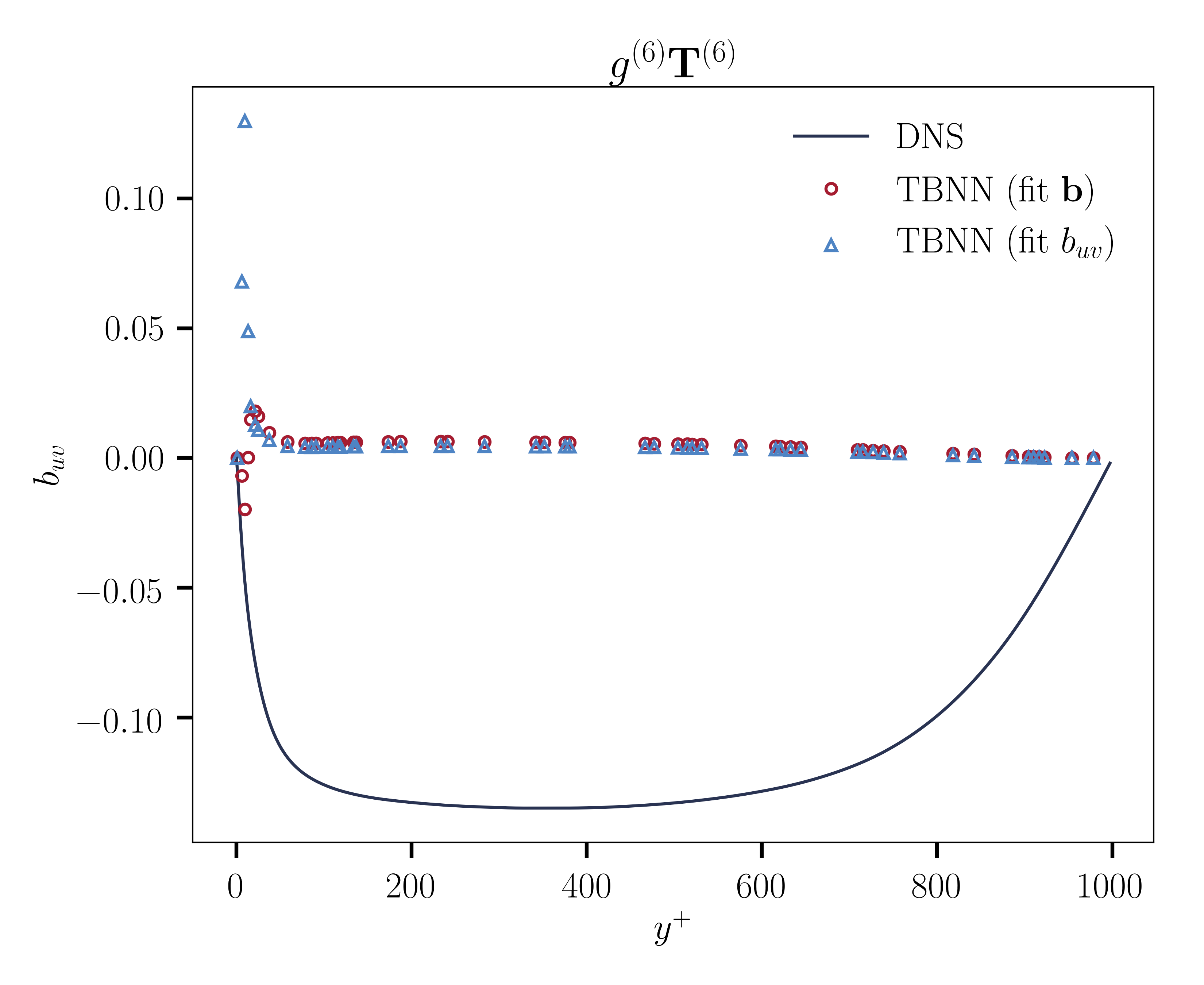}
        \caption{Predictions of term $g^{\lr{6}} \mathbf{T}^{\lr{6}}$ in the $u-v$ component.}
        \label{fig:buv_TBNN_g6T6}
    \end{subfigure}
    \caption{Predictions of $b_{uv}$ by individual active basis tensors of the GEVM.}
    \label{fig:buv_TBNN_basistensors}
\end{figure}

As discussed in section \ref{sec:channelflow}, the active components for the GEVM for the channel flow are $g^{\lr{1}} \mathbf{T}^{\lr{1}}$ and $g^{\lr{6}} \mathbf{T}^{\lr{6}}$. Figure~\ref{fig:buv_TBNN_basistensors} shows the profiles of these two terms in the $u-v$ component. 
Comparing the profiles to the overall $b_{uv}$ predictions by the TBNN models (figure~\ref{fig:buv_TBNN}), it is clear that the dominant contribution is from the linear portion $g^{\lr{1}} \mathbf{T}^{\lr{1}}$. The contribution from the nonlinear portion $g^{\lr{6}} \mathbf{T}^{\lr{6}}$ only becomes obvious in the near-wall region. This makes sense because in the bulk region the gradient $\frac{\mathrm{d}\uave}{\mathrm{d}y}$ is very small, and we know from \eqref{eq:buv} that the linear term is proportional to $\frac{k}{2\epsilon}\frac{\mathrm{d}\uave}{\mathrm{d}y}$ whereas the nonlinear term is proportional to $\lr{\frac{k}{2\epsilon}\frac{\mathrm{d}\uave}{\mathrm{d}y}}^3$. The TBNN results therefore validate the importance of the linear term and provide justification for the assumption of the linear eddy viscosity model away from the wall.

\begin{figure}[h!]
  \centering
  \includegraphics[width=0.7\textwidth]{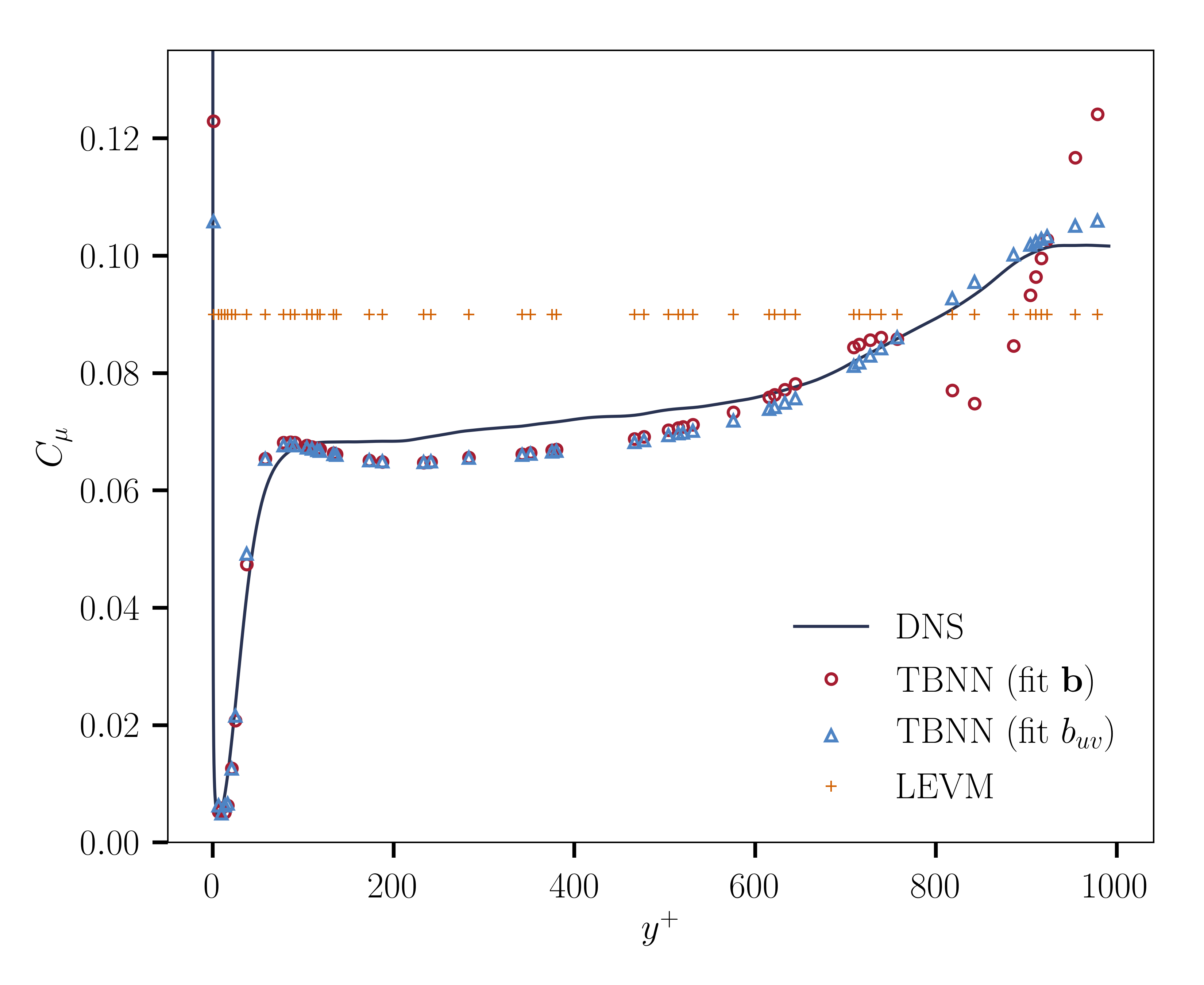}
  \caption{Comparison of $C_{\mu}$ computed from the DNS and from different models.}
  \label{fig:g-cmu}
\end{figure}

To gain a deeper insight on how the TBNN differs from the LEVM, we inspect the coefficients for different models.
Recall from section~\ref{sec:channelflow} that the coefficient $C_{\mu}$ in the LEVM~\eqref{eq:levm_normalized} corresponds to $-g^{\lr{1}} + 2 g^{\lr{6}} \alpha^2$ in the GEVM~\eqref{eq:nlev} for the channel flow. Figure~\ref{fig:g-cmu} compares $C_{\mu}$ from the TBNN and the LEVM to the true values from DNS. 
This figure clearly shows that the TBNN models are able to learn the correct value of $C_\mu$ and capture the fact that it is indeed not constant.  In particular, the TBNN model is able to match the value of $C_{\mu}$ in the near-wall region.  Figure~\ref{fig:g-cmu} also shows that the TBNN trained on $b_{uv}$ provides better predictions near the center of the channel than the TBNN trained on the full tensor, as discussed in section~\ref{sec:profiles}.  Although $g^{\lr{1}}$ and $g^{\lr{6}}$ only appear in the $u-v$ component, their values are still influenced by the other components in the GEVM.  Hence, predictions of  $g^{\lr{1}}$ and $g^{\lr{6}}$ trained on the entire tensor may be polluted by the fact that the TBNN model is trying to compensate for the GEVM deficiency near the center of the channel.

\subsection{Evolution of expansion coefficients}
The TBNN performs well at learning the components of the Reynolds stress tensor and it is also able to find the correct spatial profile for the coefficients in different regions of the flow field.  However, it is still limited by the mathematical form of the GEVM.  That is, the GEVM has an intrinsic limitation in that it requires $\nanirs$ to be identically zero in the center of the channel.  Next, we inspect if and how the TBNN model copes with this deficiency.

\begin{figure}[h!]
  \centering
  \includegraphics[width=0.7\textwidth]{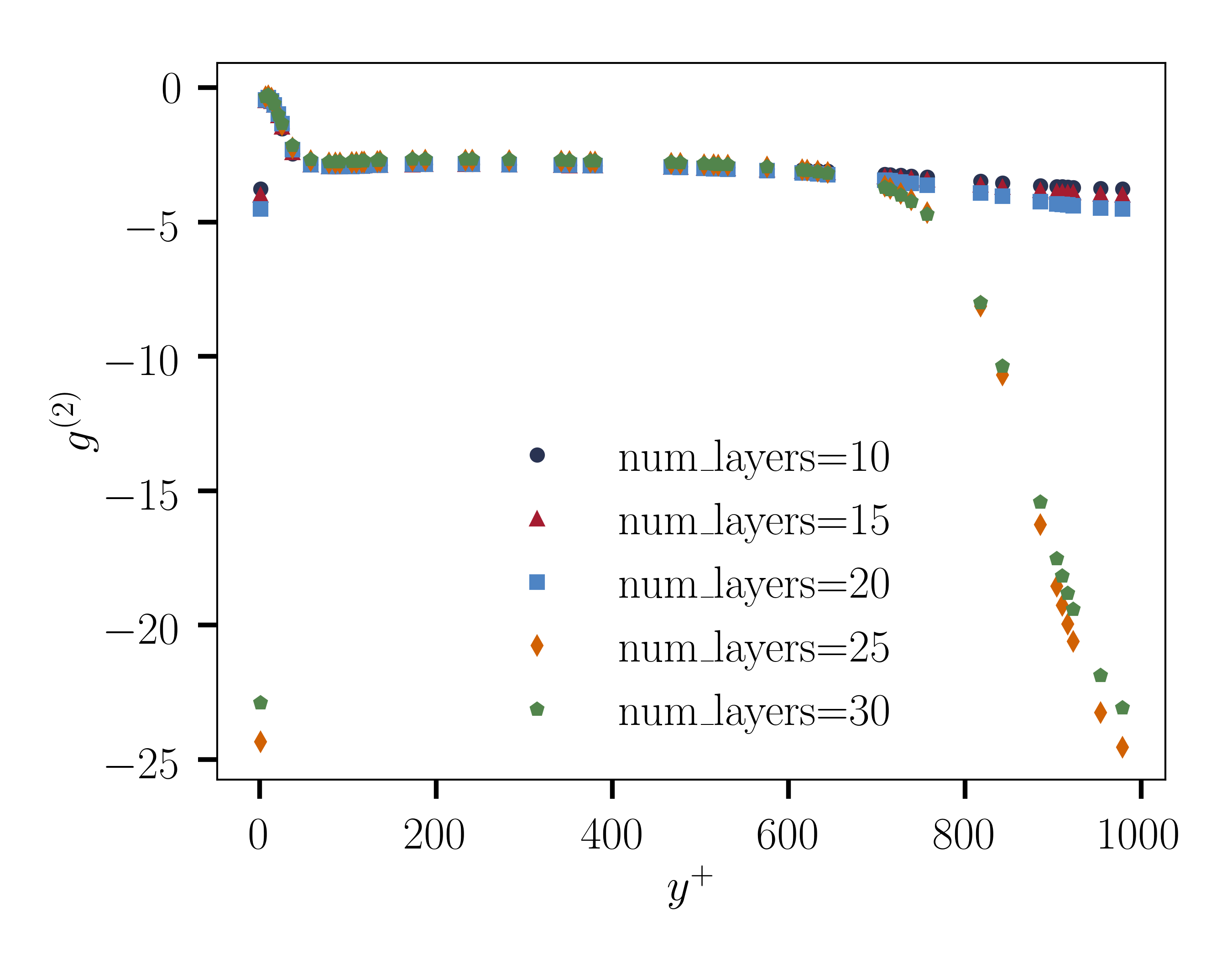}
  \caption{Profiles of the coefficient $g^{\lr{2}}$ across the channel the TBNN models with various numbers of layers.  Near the center of the channel, the magnitude of the coefficient starts to grow as the network attempts to fit the data better.}
  \label{fig:gevo}
\end{figure}

Figure~\ref{fig:gevo} shows the $g^{\lr{2}}$ profile for the TBNN models trained on the full tensor using different numbers of layers. Our best results are obtained with $25$ layers.  The $g^{\lr{2}}$ term is the first term in the expansion for $b_{uu}$ and $b_{vv}$ (see~\eqref{eq:buu}).  The DNS data indicates that $b_{uu}$ is nonzero at the center of the channel.  However, since the GEVM only depends on local velocity gradients, $b_{uu}$ is identically zero at the center of the channel and the TBNN attempts to compensate for this by making the coefficient $g^{\lr{2}}$ very large near the center of the channel.  Ultimately, this behavior has a negative impact on $b_{uu}$ away from the center, but it is interesting nonetheless that TBNN shows a ``reaction" to the flaw hardwired into its structure.

\section{Conclusions} \label{sec:conclusions}
The tensor basis neural network of~\cite{ling2016reynolds} was analyzed on a turbulent channel flow.  While previous work focused on the 
predictive capabilities of the proposed network on a variety of flow fields, the aim of the current paper was to analyze how and what the  
TBNN is actually learning for the specific case of turbulent channel flow.  We began our analysis by assessing the performance of the TBNN 
in predicting various components of the anisotropic Reynolds stress tensor and found that, unsurprisingly, the TBNN outperforms the 
classical linear as well as a quadratic eddy viscosity model.  We traced this enhanced performance to the ability of the TBNN to learn that 
the coefficient of the first term in the expansion exhibits a spatial profile, unlike the assumption of both linear and quadratic models.  
In fact, the coefficient in the first term of the expansion nearly matches the DNS results.  

We also explored the functional form of the tensor basis coefficients, namely their spatial dependence on the cross-flow coordinate $y^+$.  
We found that the TBNN makes an effort to overcome a fundamental deficiency of the general eddy viscosity model, namely a zero value of the 
Reynolds stress tensor at the center of the channel, in contrast with evidence from DNS data.  Interestingly, the TBNN attempts to 
compensate for this deficiency by making the coefficients very large near the center of the channel.  This observation suggests that the 
shortcomings of the TBNN may be addressed by an alternative architecture based on an extended tensor representation.

Several avenues for future exploration can be devised.  Although working with a single, canonical flow field was beneficial for exploring in 
close detail the mechanism by which the network learns the physics of channel flow turbulence, it is clear that deeper insight would be 
gained by analyzing further canonical flows, such as the backward facing step or a cube in crossflow.  Likewise, different neural network 
architectures might prove more effective for complicated flow fields with less symmetry than channel flow. This is certainly a major leap of 
complexity, because symmetries impose major constraints on the realizable regions of hyper-parameter space. 

Another fundamental issue encountered in the present study is non-locality.  The Reynolds stress tensor is known to be spatially and 
temporally nonlocal for general flow fields~\cite{hamba2005nonlocal, kraichnan1987eddy}.  The GEVM model used in the current work assumed 
spatial and temporal locality of the stress tensor.  Building a neural network that can account for nonlocality may offer a promising route 
forward.  For example, a recurrent neural network may be able to account for temporal nonlocality.  Additionally, kinetic models of 
turbulence based on the lattice Boltzmann equation may prove particularly well-suited for addressing nonlocality because they encode 
nonlocal effects within a local relaxation time in extended phase-space~\cite{chen2003extended}.  Indeed, by promoting the local 
relaxation to the status of a spacetime dependent field, obeying its own equation of motion, such kinetic models can in principle account 
for the strong heterogeneity which drives non-local physics.  

Turbulence modeling is a longstanding and highly demanding subject. Injecting important physical laws (e.g. conservation laws) within a 
machine learning harness shows promise to mark important strides towards the goal of improving our knowledge on the basic physics of 
turbulence.

\section{Acknowledgments}
One of the authors (SS) kindly acknowledges financial support from the European Research Council under the European Union's Horizon 2020 Framework Program (No. FP/2014-2020)/ERC Grant Agreement No. 739964 (COPMAT).

 \printbibliography

\end{document}